\documentclass[smallextended]{svjour3}       % onecolumn (second format)

\smartqed  
\usepackage{latexsym}
\usepackage{amsfonts}
\usepackage{graphicx} 
\usepackage{subfigure} 

\usepackage{amssymb}
\usepackage{stmaryrd}
\usepackage{natbib}
\usepackage{enumerate}
\usepackage{amsmath}

\usepackage[usenames]{color}

\def\BibTeX{{\rm B\kern-.05em{\sc i\kern-.025em b}\kern-.08em
    T\kern-.1667em\lower.7ex\hbox{E}\kern-.125emX}}

\begin{document}

%\title{Generalized Multiple Try Metropolis schemes}
\title{On the flexibility of the design of  Multiple Try Metropolis schemes}
\author{Luca Martino \and Jesse Read}
%Department of Signal Theory and Communications, Universidad Carlos III de Madrid.\\
%Avenida de la Universidad 30, 28911 Legan\'es, Madrid, Spain.\\
%E-mail: {\tt luca@tsc.uc3m.es, jesse@tsc.uc3m.es}}
\institute{Luca Martino \and Jesse Read \at
              Department of Signal Theory and Communications, Universidad Carlos III de Madrid \\
              Tel.:  0034-916249192\\
          %    Fax: +123-45-678910\\
              \email{luca@tsc.uc3m.es, jesse@tsc.uc3m.es}           %  \\
}
\date{Received: date / Accepted: date}

\maketitle

%\vspace{-1.5cm}
\begin{abstract}
%Markov chain Monte Carlo (MCMC) techniques are very important their relevance for statistical inference, in particular Bayesian inference.
The Multiple Try Metropolis (MTM) method is a generalization of the classical Metropolis-Hastings algorithm in which the next state
of the chain is chosen among a set of samples, according to normalized weights. In the literature,  several extensions have been proposed.
In this work, we show and remark upon the flexibility of the design of MTM-type methods, fulfilling the detailed balance condition.  
We discuss several possibilities and show different numerical results. 
%(\citep{Metropolis53} and \citep{Hastings70}) , \citep{Liu00},  \citep{Qin01},   \citep{Pandolfi10}, \citep{Casarin10}, \citep{LucaJesse11}
\end{abstract}
%\begin{keywords}
%\end{keywords}
%\IEEEpeerreviewmaketitle

%%%%%%%%%%%%%%%%%
%%%%%%%%%%%%%%%%%
\section{Introduction}
%%%%%%%%%%%%%%%%%
%%%%%%%%%%%%%%%%%
Monte Carlo methods are very useful tools for scientific and approximate computing, numerical inference and optimization \citep{Devroye86,Robert04}.
For instance, Monte Carlo methods are often necessary for the implementation of optimal Bayesian estimators so that several families of techniques have been proposed  \citep{Fitzgerald01,Gilks95bo}. %that enjoy numerous applications, from statistical physics problems \citep{Rosenbluth55,Siepmann92} to nuclear medicine applications \citep{Ljungberg98}.
 The core of the Monte Carlo approach consists of drawing random samples from a target probability density function (pdf).

%Bayesian methods have become very popular in signal processing during the past decades and, with
%them, there has been a surge of interest in the Monte Carlo techniques that are often necessary for the
%implementation of optimal a posteriori estimators [5, 7, 16, 17]. 

A very powerful class of Monte Carlo techniques is the so-called Markov Chain Monte Carlo (MCMC) algorithms \citep{Gamerman97bo,Gilks95bo,Liang10,Liu04b,Robert04}. They generate a Markov chain such that its stationary distribution coincides with the target probability density function (pdf). Typically, the only requirement is to be able to evaluate the target function, where the knowledge of the normalizing constant is usually not needed. 

The most popular MCMC method is undoubtedly the Metropolis-Hasting (MH) algorithm \citep{Hastings70, Metropolis53}. It can be applied to almost any arbitrary target distribution. However, to speed up the convergence and reduce the ``burn-in'' period, several extensions have been proposed in literature. For instance, the Multiple Try Metropolis (MTM) scheme  \citep{Liu00} where, according to certain weights, the next state of the Markov chain is selected from a set of independent samples drawn from a generic proposal density. The main advantage of MTM is that it can explore a larger portion of the sample space without a decrease of the acceptance rate. Previously, a similar methodology was proposed in the domain of molecular simulation, called  ``orientational bias Monte Carlo'' \citep[Chapter 13]{Frenkel96}, where i.i.d.\ candidates are drawn from a {\it symmetric} proposal pdf and one of these is chosen according to normalized weights directly proportional to the target pdf.

Due to its good performance and the attractive possibility to combine it with adaptive MCMC strategies \citep[Chapter 8]{Liang10}, \citep{Haario01} (for instance using different interacting or adaptive proposals at the same iteration \citep{Casarin10}), the basic formulation of the MTM has been modified and stressed in different ways.
In \citep{Pandolfi10}, the transition rule of the MTM algorithm is generalized such that the analytic form of the weights is not specified. They also study the extension of the MTM in the reversible jump framework. In \citep{Casarin10} an MTM scheme with different proposal is introduced. Different approaches with correlated candidates have been suggested in \citep{Craiu07,LucaJesse11,Qin01}.     
Some interesting theoretical results on the asymptotic behavior of different MTM strategies and some considerations on the choice of the weights are given in \citep{Bedard12}.

In all the proposed MTM schemes the number of generated candidates is fixed, differently from the delayed rejection Metropolis algorithm \citep{Mira01,Tierney99}, and the state space is not augmented defining an extended target distribution, as in other MCMC methods based on auxiliary random variables \citep{Storvik11}.

 %%%%%%%%%%%%%%%%%%%
%%%%%%%%%%%%%%%%%%%%%%%%%%%%
%%%%%%%%%%%%%%%%%%%%%%%%%%%%
In this work, we stress and remark upon the flexibility in the choice of transition rules within MTM algorithms.  First of all, we mix the approaches from \citep{Casarin10} and \citep{Pandolfi10},  building a MTM with generic weights using different proposal pdfs. Then, we present a general framework for the construction of acceptance probabilities in MTM schemes. We show this theoretically and illustrate with specific examples. Owing to this flexibility,  it is also possible to design a MTM scheme without drawing reference points \citep{RobertBlog}. Moreover, we also introduce this kind of MTM algorithm with a determinist reference points, and then discuss how this change affects its performance. 
We show that all the presented schemes fulfill the detailed balance condition and provide numerical comparisons.  
Related considerations can be found in \citep{Barker65,Brooks98,Hastings70,Peskun73,Storvik11,Tierney94,Zhang12}. 

%anFurther interesting and related considerations about the use of auxiliary variables for building acceptance probabilities within a MH approach can be found in \citep{Storvik11}.%\marginpar{las 

%CITARE STORVIK, PESKUN y igual HASTING70 sobre el tema de la flexilibilidad..  

%all proofs are provided....Moreover, we prove that our novel schemes fulfill the detailed balance condition and also provide a numerical example. 
%%%%%%%%%%%%%%%%%%%%%%%%%%%%
%%%%%%%%%%%%%%%%%%%%%%%%%%%%
%%%%%%%%%%%%%%%%%%%%%%%%%%%%

The rest of the paper is organized as follows. In Section \ref{Back} we combine the schemes in  \citep{Casarin10, Pandolfi10} describing an MTM algorithm using different proposal densities and generic weight functions. In Section \ref{NovelSect}, we explain the flexibility in the choice of the acceptance functions, satisfying the detailed balance condition. Some examples of acceptance rules are shown in Section \ref{ExampleAlfa}. Section \ref{MTMwithoutReferencepoints} introduces a MTM method without generating the reference points randomly. Numerical comparisons are given in Section \ref{ToyExample} and finally we draw conclusions in Section \ref{SectConcl}.

%%%%%%%%%%%%%%%%%%%%
%%%%%%%%%%%%%%%%%%%%
%%%%%%%%%%%%%%%%%%%%

%%%%%%%%%%%%%%%%%%%%%%%%%%%%%%%%%%%%%%%%%
%%%%%%%%%%%%%%%%%%%%%%%%%%%%%%%%%%%%%%%%%
\section{MTM algorithm with generic weights and different proposals}
%%%%%%%%%%%%%%%%%%%%%%%%%%%%%%%%%%%%%%%%%
%%%%%%%%%%%%%%%%%%%%%%%%%%%%%%%%%%%%%%%%%
\label{Back}
In the classical MH algorithm, a new possible state is drawn from the proposal pdf and the movement is accepted with a decision rule that guarantees fulfillment of the balance condition. In a multiple try approach, several (independent \citep{Liu00,Pandolfi10} or correlated \citep{LucaJesse11,Qin01})  samples are generated and from these a ``good'' one is chosen. 

In  \citep{Casarin10} the standard MTM is generalized using different proposal densities whereas  in \citep{Pandolfi10} the authors extend the standard MTM considering generic weight functions. In the following section, we recall and mix together both approaches  \citep{Casarin10, Pandolfi10} providing  an extended MTM algorithm drawing candidates from with different proposals  where the weight functions are not defined specifically, i.e., the analytic form can be chosen arbitrarily (they must be bounded and positive functions). %Moreover, we describe it in a more generic form where different proposal functions are used to draw the $N$ samples. 

%%%%%%%%%%%%%%%%%%%%%%%%%%%%%%%%%%%%%%%%%%%
\subsection{Algorithm}
%%%%%%%%%%%%%%%%%%%%%%%%%%%%%%%%%%%%%%%%%%%%
\label{MTM_Pandolfi}
Let $p_o(x)$ be the pdf that we want to draw from and $p(x)$ a function proportional to our target pdf $p_o(x)$ (i.e., $p(x)\propto p_o(x)$).  
Given a current state of the chain $x_t=x\in \mathcal{D}\subseteq \mathbb{R}$, $t\in \mathbb{N}$,  (we assume scalar values only for simplicity in the treatment), we draw $N$ independent samples each step from different proposal pdfs, i.e.,
\begin{gather}
\begin{split}
\nonumber
y_1\sim \pi_1(\cdot|x), y_2\sim \pi_2(\cdot|x),\ldots, y_N\sim \pi_N(\cdot|x).
\end{split}
\end{gather}
Therefore, we can write the joint distribution of the generated samples as
\begin{equation}
%\label{EqQ1}
\nonumber
q_N(y_{1:N}|x)=\pi_1(y_1|x)\pi_2(y_2|x)\cdots \pi_N(y_N|x).
\end{equation}
Then, a ``good'' candidate among the generated samples is chosen according to weight functions $\omega(z_1,z_2)\in \mathbb{R}^{2}\rightarrow \mathbb{R}^{+}$
(where $z_1$ and $z_2$ are generic variables) that have to be (a) bounded and (b) positive.
Given a current state $x_{t}=x$, the algorithm can be described as follows:
\begin{enumerate}
\item Draw $N$ samples $y_{1:N}=[y_1,y_2,...,y_N]$ from the joint pdf
\begin{equation}
\nonumber
q(y_{1:N}|x)=\pi_1(y_1|x)\pi_2(y_2|x)\pi_2(y_3|x)\cdots \pi_N(y_N|x),
\end{equation}
namely, draw $y_j$ from $\pi_j(\cdot|x)$, with $j=1,...,N$. 
\item Calculate the weights $\omega_j(y_{j},x)$, $j=1,...,N$, and normalize them to obtain $\bar{\omega}_j$, $j=1,...,N$.
\item Draw a $y=y_k\in\{y_1,....,y_N\}$ according to $\bar{\omega}_j$, $j=1,...,N$ and set (recall that $y_k=y$)
\begin{equation}
\label{DefWy}
W_y=\bar{\omega}_k=\frac{\omega_k(y,x)}{\sum_{j=1}^{N}\omega_j(y_{j},x)}.
\end{equation}
\item Draw other auxiliary samples (often called {\it reference points}),
\begin{equation}
\nonumber
x_{i}^{*}\sim \pi_i(\cdot|y)
\end{equation}
for $i=1,...,k-1,k+1,....,N$, and set $x_{k}^{*}= x$.
\item Compute the corresponding weights $\omega_j(x_{j}^{*},y)$, $j=1,...,N$ and set (recall that $x_{k}^{*}=x$)
\begin{equation}
\label{DefWx}
W_x=\frac{\omega_k(x,y)}{\sum_{j=1}^{N}\omega_j(x_{j}^{*},y)}.
\end{equation}
\item Let $x_{t+1}=y$ (recall that $y=y_k$) with probability
\begin{equation}
\label{alpha1trad}
\alpha(x,y)=\min\left[1,\frac{p(y)\pi_k(x|y)}{p(x)\pi_k(y|x)}\frac{W_x}{W_y}\right],
\end{equation}
otherwise set $x_{t+1}=x$ with the remaining probability $1-\alpha(x,y)$. 
\item Set $t=t+1$ and go back to the step 1.
\end{enumerate}
The kernel of the algorithm above satisfies the detailed balance condition. The proof is a special case of the development that we will present in Section \ref{SectProof}, using the probability $\alpha(x,y)$ in Eq. \eqref{alpha1trad}.

%%%%%%%%%%%%%%%%%%%%%%%%%%%%%%
\subsection{Special case: standard MTM algorithm}
\begin{figure}[htb]
\centerline{
 		\includegraphics[width=7cm]{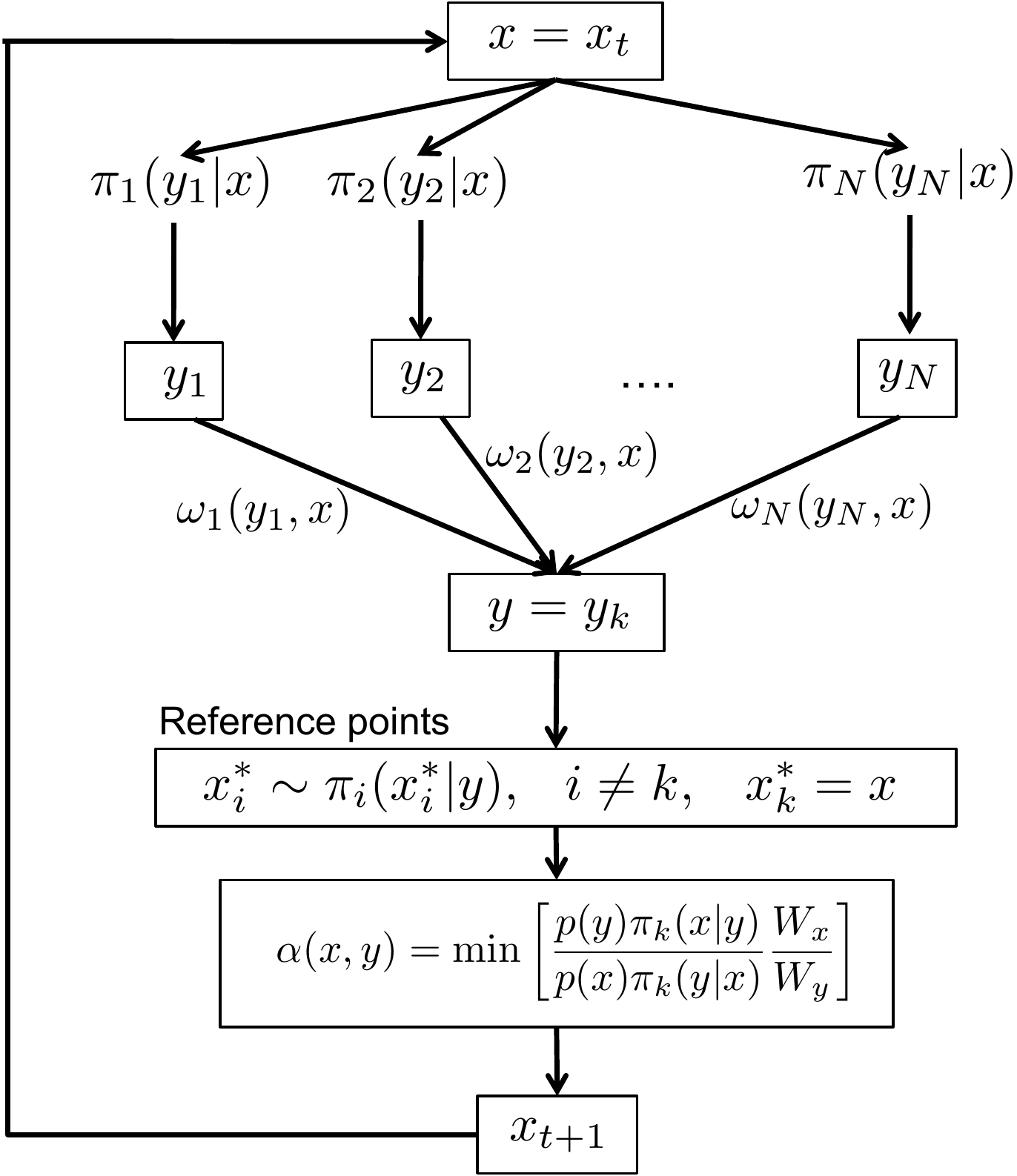} 
 		}
\caption{Sketch of the MTM algorithm with generic weights and different proposals described in Section \ref{MTM_Pandolfi}.}
\label{fig2}
\end{figure}

%%%%%%%%%%%%%%%%%%%%%%%%%%%%%%
\label{SpecialCaseStandMTM}
Choosing the weight functions with the specific analytic form
\begin{equation}
\label{Weightforms}
\omega_i(y_i,x)=p(y_i)\pi_i(x|y_i) \lambda_i(x,y_i),
\end{equation}
with $\lambda_i(x,y_i)= \lambda_i(y_i,x)$, $i=1,...,N$, we obtain the MTM  scheme proposed in \citep{Casarin10} (with different proposals). Indeed, note that the acceptance function \eqref{alpha1trad} can be also expressed as
\begin{equation}
%\label{alpha1trad_MHtrad}
\nonumber
\alpha(x,y)=\min\left[1,\frac{p(y)\pi_k(x|y)}{p(x)\pi_k(y|x)}\frac{\omega_k(x,y)}{\omega_k(y,x)}\frac{\sum_{j=1}^{N}\omega_j(y_{j},x)}{\sum_{j=1}^{N}\omega_j(x_{j}^*,y)}\right],
\end{equation}
and using the weight choice in Eq. \eqref{Weightforms} ,
\begin{equation}
%\label{alpha1trad_MHtrad2}
\nonumber
\alpha(x,y)=\min\left[1,\frac{p(y)\pi_k(x|y)}{p(x)\pi_k(y|x)}\frac{p(x)\pi_k(y|x) \lambda_k(x,y)}{p(y)\pi_k(x|y) \lambda_k(y,x)}\frac{\sum_{j=1}^{N}\omega_j(y_{j},x)}{\sum_{j=1}^{N}\omega_j(x_{j}^*,y)}\right],
\end{equation}
then it is simplified 
\begin{equation}
%\label{alpha1trad_MHtrad3}
\nonumber
\alpha(x,y)=\min\left[1,\frac{\sum_{j=1}^{N}\omega_j(y_{j},x)}{\sum_{j=1}^{N}\omega_j(x_{j}^*,y)}\right].
\end{equation}
Finally, observe that if we use just one proposal, $\pi_1(y|x)=\pi_2(y|x)=...=\pi_N(y|x)$ and  the same functions $\lambda_1(x,y)=\lambda_2(x,y)=...=\lambda_N(x,y)$, we obtain the standard formulation of the MTM \citep{Liu00}.
Figure \ref{fig2} represents a general scheme of the algorithm described in Section \ref{MTM_Pandolfi}.
%Figure \ref{fig1} illustrates the target density $p_o(x)$ (solid line) and the normalized histogram of $100,000$ samples drawn from the MTM scheme using $\alpha_{1,3}(x,y)$ and $N=10$. We can observe that the histogram approximates closely the shape of the target pdf, i.e., the Markov chain generated by the novel scheme converges to $p_o(x)$. 

%The kernel of the previous technique satisfies the detailed balance condition as shown in \citep{Liu00,Pandolfi10}.

%%%%%%%%%%%%%%%%%%%%%%%%%%%%%%
\subsection{Important observations}
%%%%%%%%%%%%%%%%%%%%%%%%%%%%%%

It is important to remark that, in order to obtain a fair comparison among the generated candidates, in the computation of the weights, it is advisable to use proposal functions with the same area below, i.e., $\int_{\mathcal{D}}\pi_1(y_1|x)dy_1=\int_{\mathcal{D}}\pi_2(y_2|x)dy_2=...=\int_{\mathcal{D}}\pi_N(y_N|x)dy_N$, for instance they can be normalized.  This is not strictly needed but recommendable.

Moreover, it is possible to show (see Section \ref{SectProof}) that the algorithm above works owing to $\alpha(x,y)$ satisfies the following equation
\begin{equation}
\label{CondGeneral}
p(x)\pi_k(y|x)W_y\alpha(x,y)=p(y)\pi_k(x|y)W_x\alpha(y,x).
\end{equation}
Note that $0 \leq W_y \leq 1$ and $0 \leq W_x \leq 1$ are probabilities and functions of $x$, $y$, the remaining points $y_i$ and $x_i^*$, then a more appropriate notation would be $W_y(y_1,...,y_k=y,...,y_N,x)$ and $W_x(x_1^*,...x_k^*=x,...,x_N^*,y)$.\footnote{Recall that $y_i$ are drawn from $\pi_i(\cdot|x)$ whereas  $x_i^*$ are drawn from $\pi_i(\cdot|y)$, $i=1,...,N$.}
However, for simplicity we maintain the notation $W_y$ and $W_x$.   
In the sequel, we suggest different acceptance functions $\alpha(x,y)$.

%%%%%%%%%%%%%%%%%%%%%%%%%
%%%%%%%%%%%%%%%%%%%%%%%%%
\section{Flexibility of the acceptance function}
%%%%%%%%%%%%%%%%%%%%%%%%%
%%%%%%%%%%%%%%%%%%%%%%%%%
\label{NovelSect}

%%We extend the previous scheme...  NOOO!!

Here, we introduce different multiple try MH approaches with generic weights functions.  Specifically 
we show how to design different suitable acceptance functions $\alpha(x,y)$ fulfilling the detailed balance condition. 
Indeed, it is possible to choose functions $\alpha(x,y)$ with the form  
\begin{equation}
\nonumber
\alpha(x,y)=\beta(x,y)\gamma(x,y|{\bf x}^{*}_{-k},{\bf y}_{-k}),
\end{equation}
where 
\begin{enumerate}
\item $\beta(x,y)$ is such that  
\begin{equation}
\label{condBeta}
p(x)\pi_k(y|x)\beta(x,y)= p(y)\pi_k(x|y)\beta(y,x),\mbox{  }\mbox{  }\mbox{  } \forall k\in \{1,...,N\}, 
\end{equation}
\item $\gamma(x,y|{\bf x}^{*}_{-k},{\bf y}_{-k})$ satisfies 
\begin{equation}
\label{condGamma}
W_y\gamma(x,y|{\bf x}^{*}_{-k},{\bf y}_{-k})=W_x\gamma(y,x|{\bf y}_{-k},{\bf x}^{*}_{-k}),
\end{equation}
where ${\bf x}^{*}_{-k}= [x^{*}_1,...x^{*}_{k-1},x^{*}_{k+1},...,x^{*}_{N}]$ and ${\bf y}_{-k}= [y_1,...y_{k-1},y_{k+1},...,y_{N}]$.
\item Finally we need 
\begin{equation}
\label{Alfamenor1}
0 \leq \alpha(x,y)\leq 1.
\end{equation}
\end{enumerate}
If the Eqs. (\ref{condBeta}) and (\ref{condGamma}) are jointly fulfilled then the condition (\ref{CondGeneral}) also holds, i.e., the equation
$$p(x)\pi_k(y|x)W_y\alpha(x,y)=p(y)\pi_k(x|y)W_x\alpha(y,x) $$
is satisfied. Equation (\ref{Alfamenor1}) can be easily obtained choosing separately $0 \leq \beta(x,y) \leq 1$ and $0 \leq \gamma(x,y|{\bf x}^{*}_{-k},{\bf y}_{-k}) \leq 1$. Moreover, in this case, Eq. (\ref{condBeta}) is exactly the balance condition of the standard MH algorithm, then we can choose any acceptance functions suitable for the standard MH algorithm as function $\beta(x,y)$. Similar considerations can be used to design suitable functions $\gamma(x,y|{\bf x}^{*}_{-k},{\bf y}_{-k})$. Some examples are provided in Section \ref{ExampleAlfa}.

%%%%%%%%%%%%%%%%
%%%%%%%%%%%%%%%%
\subsection{Algorithm}
%%%%%%%%%%%%%%%%
%%%%%%%%%%%%%%%%
The novel scheme can be summarized as follows:
\begin{enumerate}
\item Draw $N$ samples from the proposal pdfs $y_j \sim \pi_j(\cdot|x)$, with $j=1,...,N$. 
\item Calculate the weights $\omega_j(y_{j},x)$, $j=1,...,N$, and normalize them to obtain $\bar{\omega}_j$, $j=1,...,N$.
\item Draw a $y=y_k\in\{y_1,....,y_N\}$ according to $\bar{\omega}_j$, $j=1,...,N$ and set (recall that $y_k=y$)
$$W_y=\bar{\omega}_k=\frac{\omega_k(y,x)}{\sum_{j=1}^{N}\omega_j(y_{j},x)}.$$
\item Draw other auxiliary samples $x_{i}^{*}\sim \pi_i(\cdot|y)$ for $i=1,...,k-1,k+1,...,N$, and set $x_{k}^{*}= x$.
\item Compute the corresponding weights $\omega_j(x_{j}^{*},y)$, $j=1,...,N$ and set (recall that $x_{k}^{*}=x$)
$$W_x=\frac{\omega_k(x,y)}{\sum_{j=1}^{N}\omega_j(x_{j}^{*},y)}.$$
\item Let $x_{t+1}=y$ (recall that $y=y_k$) with probability
\begin{equation}
%\label{alphaNEW}
\nonumber
\alpha(x,y)=\beta(x,y)\gamma(x,y|{\bf x}^{*}_{-k},{\bf y}_{-k}),
\end{equation}
where 
$$p(x)\pi_k(y|x)\beta(x,y)= p(y)\pi_k(x|y)\beta(y,x)$$
and 
$$W_y\gamma(x,y|{\bf x}^{*}_{-k},{\bf y}_{-k})=W_x\gamma(y,x|{\bf y}_{-k},{\bf x}^{*}_{-k}).$$
Otherwise set $x_{t+1}=x$ with the remaining probability $1-\alpha(x,y)$. 
\item Set $t=t+1$ and go back to the step 1.
\end{enumerate}

%%%% 

% abajo lo estoy haciendo con diferentes pesos y diferentes proposal functions....
%%%%

%%%%%%%%%%%%%%%%%%%%%%%
%%%%%%%%%%%%%%%%%%%%%%%
\subsection{Balance condition}
%%%%%%%%%%%%%%%%%%%%%%%
%%%%%%%%%%%%%%%%%%%%%%%
\label{SectProof}

To guarantee that a Markov chain generated by an MCMC method converges to the target distribution $p_o(x)\propto p(x)$, we can prove that the kernel $A(y|x)$ of the corresponding algorithm (probability of accepting a generated sample $y$ given the previous state value $x$) fulfills the following detailed balance condition\footnote{Note that the balance condition is a sufficient but not necessary condition. Namely, the detailed balance ensures invariance. The converse is not true. Markov chains that satisfy the detailed balance condition are called {\it reversible}.} \citep{Liu04b,Robert04}  
\begin{equation}
\nonumber
p(x)A(y|x)=p(y)A(x|y).
\end{equation}
First of all, we need to write down the kernel $A(y|x)$. We consider $x\neq y$, since the case $x=y$ is trivial (indeed, in this case $A(y|x)$ is proportional to a delta function $\delta(y-x)$ \citep{Liu04b,Robert04}). 
The kernel (for $x\neq y$) can be expressed as 
 \begin{equation}
 \nonumber
A(y=y_k|x)=\sum_{i=1}^{N} h(y=y_k|x,k=i),
\end{equation}
where $h(y=y_k|x,k=i)$ is the probability of accepting  the new state $x_{t+1}=y_k$ given the previous one $x_{t}=x$, when the chosen sample $y_k$ is the $i$-th candidate, i.e., when $y_k=y_i$. However, since the $y_i$ are exchangeable, for symmetry we have $h(y=y_k|x,i)=h(y=y_k|x,j)$ $\forall i,j\in\{1,...,N\}$. Hence, we can also write 
 \begin{equation}
\nonumber
A(y=y_k|x)=N \cdot h(y=y_k|x,k),
\end{equation}
where $k\in \{1,...,N\}$ and we recall $N$ is the total number of proposed candidates $y_i$. Then, we need to show that 
$$p(x)h(y|x,k)=p(y)h(x|y,k),$$ 
for a generic $k\in\{1,...,N\}$. Following each step of the algorithm above, we can write 
\begin{gather}
%\small
\nonumber
\begin{split}
 p(x)&h(y=y_k|x,k)= \\
p(x) \int_{\mathcal{D}}&\cdots\int_{\mathcal{D}} \left[\prod_{j=1}^{N}\pi_j(y_j|x)\right]\frac{\omega_k(y,x)}{\sum_{i=1}^{N}\omega_i(y_{i},x)} \left[\prod_{j=1; j\neq k}^{N}\pi_{j}(x^{*}_j|y)\right]\cdot \\
 & \underbrace{\beta(x,y)\gamma(x,y|{\bf x}^{*}_{-k},{\bf y}_{-k})}_{\alpha(x,y)} \mbox{ } dy_{1:k-1}dy_{k+1:N}dx^{*}_{1:k-1}dx^{*}_{k+1:N}.  \\ 
% &\mbox{ }\mbox{ }\mbox{ }\mbox{ }\mbox{ }\mbox{ }\mbox{ }\mbox{ }\mbox{ }\mbox{ }\mbox{ }\mbox{ }\mbox{ }\mbox{ }\mbox{ }\mbox{ }\mbox{ }\mbox{ }\mbox{ }\mbox{ }\cdot  \beta(x,y)\gamma(x,y|{\bf x}^{*}_{-k},{\bf y}_{-k}) dy_{1:k-1}dy_{k+1:N}dx^{*}_{2:N}.
\end{split}
\end{gather}
Note that each factor inside the integral corresponds to a step of the method described in the previous section. The integral is over all auxiliary variables. Since we consider $y=y_k$ and recalling the definition of $W_y$ in Eq. (\ref{DefWy}), we can rewrite the expression in this way 
\begin{gather}
%\small
\nonumber
\begin{split}
 p(x)&h(y|x,k)=\\
 p(x)& \int_{\mathcal{D}}\cdots\int_{\mathcal{D}} \pi_k(y|x)\left[\prod_{j=1,j\neq k}^{N}\pi_j(y_j|x)\right] W_y\left[\prod_{j=1; j\neq k}^{N}\pi_{j}(x^{*}_j|y)\right] \cdot \\
&\cdot \beta(x,y)\gamma(x,y|{\bf x}^{*}_{-k},{\bf y}_{-k}) \mbox{ } dy_{1:k-1}dy_{k+1:N}dx^{*}_{1:k-1}dx^{*}_{k+1:N}.  \\ 
% &\mbox{ }\mbox{ }\mbox{ }\mbox{ }\mbox{ }\mbox{ }\mbox{ }\mbox{ }\mbox{ }\mbox{ }\mbox{ }\mbox{ }\mbox{ }\mbox{ }\mbox{ }\mbox{ }\mbox{ }\mbox{ }\mbox{ }\mbox{ }\cdot  \beta(x,y)\gamma(x,y|{\bf x}^{*}_{-k},{\bf y}_{-k}) dy_{1:k-1}dy_{k+1:N}dx^{*}_{2:N}.
\end{split}
\end{gather}
and we only arrange it, obtaining 
\begin{gather}
\label{FinalEqProof}
%\small
\begin{split}
 p(x)h(y|x,k)=&\\
 \int_{\mathcal{D}}&\cdots\int_{\mathcal{D}} \left[\prod_{j=1,j\neq k}^{N}\pi_j(y_j|x)\right] \left[\prod_{j=1;j\neq k}^{N}\pi_{j}(x^{*}_j|y)\right] \cdot \\
&\cdot p(x)\pi_k(y|x)\beta(x,y)\cdot W_y\gamma(x,y|{\bf x}^{*}_{-k},{\bf y}_{-k}) \mbox{ } d{\bf y}_{-k}d{\bf x^{*}}_{-k}.  \\ 
% &\mbox{ }\mbox{ }\mbox{ }\mbox{ }\mbox{ }\mbox{ }\mbox{ }\mbox{ }\mbox{ }\mbox{ }\mbox{ }\mbox{ }\mbox{ }\mbox{ }\mbox{ }\mbox{ }\mbox{ }\mbox{ }\mbox{ }\mbox{ }\cdot  \beta(x,y)\gamma(x,y|{\bf x}^{*}_{-k},{\bf y}_{-k}) dy_{1:k-1}dy_{k+1:N}dx^{*}_{2:N}.
\end{split}
\end{gather}
Therefore, since we assume (see Eqs. (\ref{condBeta}) and (\ref{condGamma}))  
$$p(x)\pi_k(y|x)\beta(x,y)=p(y)\pi_k(x|y)\beta(y,x),$$
and 
$$W_y\gamma(x,y|{\bf x}^{*}_{-k},{\bf y}_{-k})=W_x\gamma(y,x|{\bf y}_{-k},{\bf x}^{*}_{-k}),$$
it is straightforward that the expression in Eq. (\ref{FinalEqProof}) is symmetric in $x$ and $y$. Indeed, we can exchange the notations of $x$ and $y$, and  $x^{*}_i$ and $y_j$, respectively, and the expression does not vary. Then we can write
\begin{equation}
\nonumber
p(x)h(y|x,k)=p(y)h(x|y,k).
\end{equation}
Since we have assumed a generic $k$ and $A(y=y_k|x)=h(y=y_k|x,k)$, it possible to assert that  
\begin{equation}
\nonumber
p(x)A(y|x)=p(y)A(x|y),
\end{equation}   
that is the balance condition. Therefore, the Markov chain generated by the algorithm, described in the previous section, converges to our target pdf.

%%%%%%%%%%%%%%%%%%%%%%%%%%%%
%%%%%%%%%%%%%%%%%%%%%%%%%%%%
\section{Examples of functions $\alpha(x,y)$}
%%%%%%%%%%%%%%%%%%%%%%%%%%%%
%%%%%%%%%%%%%%%%%%%%%%%%%%%%
\label{ExampleAlfa}

In this section, we provide some suitable acceptance functions $\alpha(x,y)=\mathcal{D}\times \mathcal{D}\rightarrow [0,1]$,
that satisfies the condition (\ref{CondGeneral}).
The easiest way is to obtain $\alpha(x,y)$ is to design separately suitable functions $0 \leq \beta(x,y) \leq 1$ and $0 \leq \gamma(x,y|{\bf x}^{*}_{-k},{\bf y}_{-k}) \leq 1$. %Indeed, in this case it is possible to choose the two functions $\beta(x,y)$ and $\gamma(x,y|{\bf x}^{*}_{-k},{\bf y}_{-k})$ separately.

%%%%%%%%%%%%%%%%%%%%%%%%%%%%
\subsection{Possible choices of $\beta(x,y)$}
%%%%%%%%%%%%%%%%%%%%%%%%%%%%
%%%%% TI MANCANO DELLE REFERENCE %%%%%%
\label{SectBeta}

To design a function $\beta(x,y)$ such that $0 \leq \beta(x,y) \leq 1$ and 
$$p(x)\pi_k(y|x)\beta(x,y)= p(y)\pi_k(x|y)\beta(y,x),$$
we can choose any acceptance rule suitable for the standard MH algorithm \citep{Barker65,Hastings70}. Hence, for instance, we can choose the classical acceptance rule of the  MH algorithm, i.e.,
\begin{equation}
\label{Betaclassical}
\beta_1(x,y)=\min\left[1,  \frac{p(y)\pi_k(x|y)}{p(x)\pi_k(y|x)}\right].
\end{equation}
Other possibilities are summarized in Table \ref{Tabla1} where $\lambda(x,y)$ is a symmetric non-negative function (i.e., $\lambda(x,y)\geq 0$ and  $\lambda(x,y)=\lambda(y,x)$ for all  $(x,y)\in \mathcal{D}\times \mathcal{D}$) such that 
$0 \leq \beta(x,y) \leq 1$.
\begin{table}[!hbt]
\begin{center}
\caption{Example of suitable functions $\beta(x,y)$}
\label{Tabla1}
\begin{tabular}{|c|c|} 
\hline
\normalsize Functions $\beta(x,y)$ &  \normalsize References\\ 
\hline
\hline
%&  \\
%\vspace{+0.3cm}
\normalsize
$\beta_1(x,y)=\min\left[1,  \frac{p(y)\pi_k(x|y)}{p(x)\pi_k(y|x)}\right]$ &  \citep{Hastings70,Metropolis53}   \\
%&  \\
%$\mbox{  }$ &$\mbox{  }$ & $\mbox{  }$  \\
\hline
\normalsize $\beta_2(x,y)=\frac{p(y)\pi_k(x|y)}{p(x)\pi_k(y|x)+p(y)\pi_k(x|y)}$ & \citep{Barker65}    \\
%&  \\
\hline
\normalsize $\beta_3(x,y)=\frac{\lambda(x,y)}{1+ \frac{p(x)\pi_k(y|x)}{p(y)\pi_k(x|y)}}$ & \citep{Hastings70}    \\
\hline
%&  \\
\normalsize $\beta_4(x,y)=\frac{p(y)\pi_k(x|y)}{\lambda(x,y)}$ & \citep{Liu04b,Robert04}\\
\hline
\normalsize
$\beta_5(x,y)=\frac{\lambda(x,y)}{p(x)\pi_k(y|x)}$ &  \citep{Liu04b,Robert04}\\
\hline
\normalsize
$\beta_6(x,y)=\frac{p(y)\lambda(x,y)}{\pi_k(y|x)}$ & \citep[Chapter 5]{Liu04b}\\
\hline
\normalsize
$\beta_7(x,y)=\frac{\pi_k(x|y)\lambda(x,y)}{p(x)}$ & \citep[Chapter 5]{Liu04b} \\
\hline
\end{tabular}
\end{center}
\end{table}
%It is also possible to use the acceptance function proposed by Barker, 1965 \cite{Barker65}  
%\begin{equation}
%\label{BetaBaker}
%\beta_B(x,y)=\frac{p(y)\pi_k(x|y)}{p(x)\pi_k(y|x)+p(y)\pi_k(x|y)},
%\end{equation}
%or by Hasting, 1970 \cite{Hastings70}
%\begin{equation}
%\label{BetaHasting}
%\beta_H(x,y)=\frac{\lambda(x,y)}{1+ \frac{p(x)\pi_k(y|x)}{p(y)\pi_k(x|y)}},
%\end{equation}
%where is a non-negative symmetric function, $\lambda(x,y)\geq 0$, $\lambda(x,y)=\lambda(y,x)$ and such that $0 \leq \beta_H(x,y) \leq 1$. Other possibilities are the acceptance rules proposed by Stein  
%\begin{equation}
%%\label{Stein}
%\beta_{SLuca}(x,y)=\frac{p(y)\pi_k(x|y) }{\lambda(x,y)},
%\end{equation}
%\begin{equation}
%%\label{Stein}
%\beta_{S2}(x,y)=\frac{\lambda(x,y)}{p(x)\pi_k(y|x)},
%\end{equation}
%\begin{equation}
%%\label{Stein}
%\beta_{S2}(x,y)=\frac{p(y)\lambda(x,y)}{\pi_k(y|x)},
%\end{equation}
%\begin{equation}
%%\label{Stein}
%\beta_{S2}(x,y)=\frac{\pi_k(y|x)\lambda(x,y)}{p(x)},
%\end{equation}

Moreover, defining 
\begin{equation}
\nonumber
R(x,y)=  \frac{p(y)\pi_k(x|y)}{p(x)\pi_k(y|x)},
\end{equation}
and considering a function $F(\vartheta): \mathbb{R}^+\rightarrow [0,1]$ such that 
\begin{equation}
\nonumber
F(\vartheta)=\vartheta F(1/\vartheta),
\end{equation}
then it is possible to define a general acceptance function \citep{Gamerman97bo,Gilks95bo}
\begin{equation}
%\label{BetaGeneral}
\nonumber
\beta_g(x,y)=(F\circ R)(x,y)=F(R(x,y)).
\end{equation}
%where the subscript
For instance, if $F(\vartheta)=\min[1, \vartheta]$ we obtain Eq. (\ref{Betaclassical}) and if $F(\vartheta)=\frac{\vartheta}{1+\vartheta}$ we find $\beta_2$ or $\beta_3$ with $\lambda(x,y)=1$ (see Table \ref{Tabla1}).
In \citep{Peskun73} there is a comparison of different acceptance functions in a standard MH algorithm.

%%%%%%%%%%%%%%%%%%%%%%%%%%%%%%%%%%%%%
\subsection{Possible choices of $\gamma(x,y|{\bf x}^{*}_{-k},{\bf y}_{-k})$}
%%%%%%%%%%%%%%%%%%%%%%%%%%%%%%%%%%%%%
\label{SectGamma}
In this section, we provide some examples of suitable function $\gamma(x,y|{\bf x}^{*}_{-k},{\bf y}_{-k})$. 
We need functions $\gamma(x,y|{\bf x}^{*}_{-k},{\bf y}_{-k})$ such that 
\begin{equation}
\label{condGamma2}
W_y\gamma(x,y|{\bf x}^{*}_{-k},{\bf y}_{-k})=W_x\gamma(y,x|{\bf y}_{-k},{\bf x}^{*}_{-k}),
\end{equation}
where
$$W_y=\frac{\omega_k(y,x)}{\sum_{j=1}^{N}\omega_j(y_{j},x)},\mbox{ }\mbox{ }\mbox{ and } \mbox{ } W_x=\frac{\omega_k(x,y)}{\sum_{j=1}^{N}\omega_j(x^{*}_{j},y)}.$$
 Therefore, for instance, it is possible to choose
\begin{equation}
\nonumber
\gamma_1(x,y|{\bf x}^{*}_{-k},{\bf y}_{-k})=W_x.
\end{equation}
Indeed, in this case $\gamma(y,x|{\bf y}_{-k},{\bf x}^{*}_{-k})=W_y$ and the condition (\ref{condGamma2}) is satisfied ($W_yW_x=W_xW_y$). 
Another possibility is to define
\begin{equation}
\nonumber
\gamma_2(x,y|{\bf x}^{*}_{-k},{\bf y}_{-k})=\frac{W_x}{W_x+W_y},
\end{equation}
or 
\begin{equation}
\nonumber
\gamma_3(x,y|{\bf x}^{*}_{-k},{\bf y}_{-k})=\min\left[1,\frac{W_x}{W_y} \right].
\end{equation} 
\section{MTM without drawing reference points}
%%%%%%%%%%%%%%%%%%%%%%%%%%%%
\label{MTMwithoutReferencepoints}

The previous considerations also suggest how it is possible to design a MTM that avoids sampling the reference points ${\bf x}^*_{-k}$. For some authors generating the reference samples is considered a drawback of the MTM schemes, since $N-1$ samples are {\it only} drawn to fulfill the balance condition \citep{RobertBlog}. To avoid this step, the MTM method in Section \ref{MTM_Pandolfi} can be modified as follows:
\begin{enumerate}
\item Given a current state $x_{t}=x$, draw $N$ samples $y_{1:N}=[y_1,y_2,...,y_N]$ from the joint pdf
\begin{equation}
\nonumber
q(y_{1:N}|x)=\pi_1(y_1|x)\pi_2(y_2|x)\pi_2(y_3|x)\cdots \pi_N(y_N|x),
\end{equation}
namely, draw $y_j$ from $\pi_j(\cdot|x)$, with $j=1,...,N$. 
\item Calculate the weights $\omega_j(y_{j},x)$, $j=1,...,N$, and normalize them to obtain $\bar{\omega}_j$, $j=1,...,N$.
\item Draw a $y=y_k\in\{y_1,....,y_N\}$ according to $\bar{\omega}_j$, $j=1,...,N$ and set 
\begin{equation}
\label{DefWyCazzo}
W_y=\bar{\omega}_k=\frac{\omega_k(y,x)}{\sum_{j=1}^{N}\omega_j(y_{j},x)}.
\end{equation}
\item Set $x_{i}^{*}= y_i$ for $i=1,...,k-1,k+1,....,N$, and set $x_{k}^{*}= x$.
\item Compute the corresponding weights $\omega_j(x_{j}^{*},y)$, $j=1,...,N$ and (recalling $x_k*=x$) set 
\begin{equation}
\label{DefWxCazzo}
W_x=\frac{\omega_k(x,y)}{\sum_{j=1}^{N}\omega_j(x_{j}^{*},y)}.
\end{equation}
\item Let $x_{t+1}=y$ (recall that $y=y_k$) with probability
\begin{equation}
\label{alphaWithout}
\alpha(x,y)=\min\left[1,\frac{p(y)\prod_{i=1}^{N}\pi_i(x_i^*|y)}{p(x)\prod_{i=1}^{N} \pi_i(y_i|x)}\frac{W_x}{W_y}\right],
\end{equation}
otherwise set $x_{t+1}=x$ with the remaining probability $1-\alpha(x,y)$. 
\item Set $t=t+1$ and go back to the step 1.
\end{enumerate}
The differences w.r.t. the standard MTM method are contained in the steps 4 and 6. In this case the vectors ${\bf y}=[y_1,...,y_k=y,....,y_N]$  and ${\bf x}^*=[x^*_1=y_1,...,x^*_k=x,....,x^*_N=y_N]$ differ only in the position $k$, i.e., ${\bf y}_{-k}={\bf x}_{-k}^*$. Hence, note that $\alpha(x,y)$ can be expressed as
\begin{equation}
\label{alphaWithout2}
\alpha(x,y)=\min\left[1,\frac{p(y)\pi_k(x|y)}{p(x)\pi_k(y|x)}\frac{\prod_{i\neq k}^{N}\pi_i(y_i|y)}{\prod_{i\neq k}^{N} \pi_i(y_i|x)}\frac{W_x}{W_y}\right].
\end{equation}
However, although this scheme satisfies the balance condition as we show below, observing the expression of $\alpha$, a drawback could seem evident: since the samples $y_{1:N}$ are drawn from $\pi_i(\cdot|x)$, $i=1,..,N$, the product $\prod_{i\neq k}^{N} \pi_i(y_i|x)$ would be ``often'' greater then $\prod_{i\neq k}^{N} \pi_i(y_i|y)$. That is to say, $x$ is more ``likely'' than $y$ given the ``observations'' $y_i$,  $i\neq k$. Therefore, $\alpha(x,y)$ would be ``often'' less than $1$ so that accepting a jump becomes ``rare''\footnote{However, it is important to remark that high acceptance rates are not a suitable indicator of good performance since, in general, the best acceptance rate is different from $1$ \citep{Roberts97}.}. This issue would increase with $N \rightarrow +\infty$. 
However, the numerical simulations (see Section \ref{ToyExample}) show that the probability $\alpha(x,y)$ first surprisingly increases for small values of $N$ (owing to the factor $\frac{W_x}{W_y}$) and then decreases with $N\rightarrow +\infty$ as expected. Moreover the performance generally gets worse with $N\rightarrow +\infty$. Hence this scheme appears, in general, useless. 
These considerations above explain as, in the standard MTM version \citep{Liu00}, the authors introduce the idea of randomly generating the reference points $x_i^*$. 
However, there is an important special case that we show in Section  \ref{indprop}. 

%%%%%%%%%%%%%%%%%%%%%
\subsection{Balance condition}
%%%%%%%%%%%%%%%%%%%%%
Again we must check that the detailed balance condition $p(x)A(y|x)=p(y)A(x|y)$ is fulfilled. 
The kernel $A(y|x)$ (for $x\neq y$) can be expressed, also in this case, as 
$A(y=y_k|x)=N \cdot h(y=y_k|x,k)$,
where $k\in \{1,...,N\}$ and $N$ is the total number of proposed candidates $y_i$. Then, finally we have to show that 
$$p(x)h(y|x,k)=p(y)h(x|y,k),$$ 
for a generic $k\in\{1,...,N\}$. Following each step of the MTM algorithm without reference point, we can write 
\begin{gather}
\small
\nonumber
\begin{split}
 p(x)h(y|x,k)=p(x) \int_{\mathcal{D}}\cdots\int_{\mathcal{D}} \left[\prod_{i=1}^{N}\pi_i(y_i|x)\right] W_y \min&\left[1,\frac{p(y)\prod_{i=1}^{N}\pi_i(x_i^*|y)}{p(x)\prod_{i=1}^{N} \pi_i(y_i|x)}\frac{W_x}{W_y}\right]  \\
 & \mbox{ } dy_{1:k-1}dy_{k+1:N}dx^{*}_{1:k-1}dx^{*}_{k+1:N}.  \\ 
% &\mbox{ }\mbox{ }\mbox{ }\mbox{ }\mbox{ }\mbox{ }\mbox{ }\mbox{ }\mbox{ }\mbox{ }\mbox{ }\mbox{ }\mbox{ }\mbox{ }\mbox{ }\mbox{ }\mbox{ }\mbox{ }\mbox{ }\mbox{ }\cdot  \beta(x,y)\gamma(x,y|{\bf x}^{*}_{-k},{\bf y}_{-k}) dy_{1:k-1}dy_{k+1:N}dx^{*}_{2:N}.
\end{split}
\end{gather}
The integral is over all auxiliary variables. Just  by rearranging, we obtain 
\begin{gather}
\label{FinalEqProof_2}
\small
\begin{split}
 &p(x)h(y|x,k)=\int_{\mathcal{D}}\cdots\int_{\mathcal{D}}  \\
 &\min\left[p(x)\prod_{i=1}^{N}\pi_i(y_i|x) W_y,p(y)\prod_{i=1}^{N}\pi_i(x_i^*|y) W_x\right] dy_{1:k-1}dy_{k+1:N}dx^{*}_{1:k-1}dx^{*}_{k+1:N}.  \\ 
\end{split}
\end{gather}
Recalling that $x_j^*=y_j$ for $j=1,..,k-1,k+1,..,N$, $x_k^*=x$ and $y_k=y$, the Eq. \eqref{FinalEqProof_2} can be rewritten as
\begin{gather}
%\label{FinalEqProof_3}
\small
\nonumber
\begin{split}
 p(x)&h(y|x,k)=\int_{\mathcal{D}}\cdots\int_{\mathcal{D}} \\
 &\min\left[p(x)\pi_k(y|x) \prod_{i\neq k}^{N}\pi_i(y_i|x) W_y,p(y)\pi_k(x|y) \prod_{i\neq k}^{N}\pi_i(y_i|y) W_x\right] \mbox{ }  dy_{1:k-1}dy_{k+1:N}.  \\ 
\end{split}
\end{gather}
Therefore  it is straightforward to see that we can exchange $x$ and $y$ without varying the expression above (see also Eq. \eqref{DefWyCazzo} and \eqref{DefWxCazzo}), then $p(x)h(y|x,k)=p(y)h(x|y,k)$ and the balance condition $p(x)A(y|x)=p(y)A(x|y)$ is satisfied.

\subsection{Independent proposal pdfs}
%%%%%%%%%%%%%%%%%%%%%%%
\label{indprop}
If the proposal pdfs are chosen as independent densities, i.e., $\pi_1(y_1|x)=\pi_1(y_1)$, $\pi_2(y_2|x)=\pi_2(y_2)$... $\pi_N(y_N|x)=\pi_N(y_N)$, the algorithm is  simplified. 
Indeed, the $\alpha(x,y)$ probability in Eq. \eqref{alphaWithout2}, i.e.,
%\begin{equation}
%\label{alphaWithout}
$$\alpha(x,y)=\min\left[1,\frac{p(y)\pi_k(x|y)}{p(x)\pi_k(y|x)}\frac{\prod_{i\neq k}^{N}\pi_i(y_i|y)}{\prod_{i\neq k}^{N} \pi_i(y_i|x)}\frac{W_x}{W_y}\right],$$
%\end{equation}
now it can be rewritten as 
\begin{equation}
\small
%\label{alphaWithout3}
\nonumber
\alpha(x,y)=\min\left[1,\frac{p(y)\pi_k(x)\prod_{i\neq k}^{N}\pi_i(y_i)}{p(x)\pi_k(y)\prod_{i\neq k}^{N} \pi_i(y_i)}\frac{W_x}{W_y}\right]=\min\left[1,\frac{p(y)\pi_k(x)}{p(x)\pi_k(y)}\frac{W_x}{W_y}\right].
\end{equation}
Observe that it is exactly the probability $\alpha(x,y)$ obtained in Eq. \eqref{alpha1trad} using independent proposals. Therefore, here, the conclusion is different from the general case: it is not necessary to draw reference points when independent proposal densities are used. It is necessary just to set deterministically $x_{i}^{*}= y_i$ for $i=1,...,k-1,k+1,....,N$, and set $x_{k}^{*}= x$.
This special case, when the weights are chosen as in Section \ref{SpecialCaseStandMTM},  is also discussed in \citep[Chapter 5]{Liu04b}.

\begin{figure}[htb]
\centerline{
 		\includegraphics[width=5.5cm]{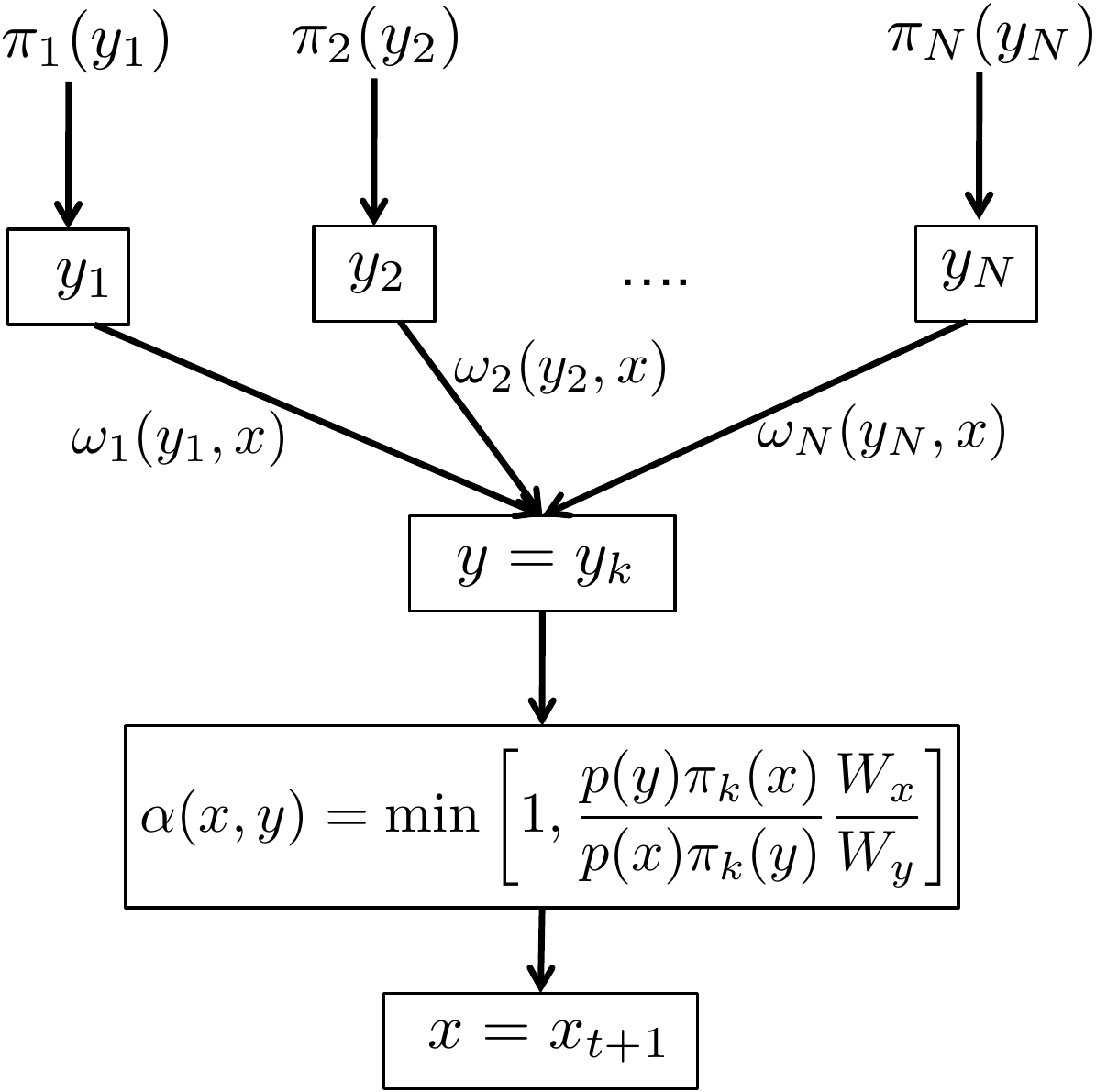}
		%\subfigure[]{\includegraphics[width=5.7cm]{FigGenScheme3.pdf}}		 
 		}
\caption{Scheme of MTM algorithm with generic weights and different independent proposal pdfs.}
\label{fig3}
\end{figure}

\begin{figure}[htb]
\centerline{
 		\subfigure[]{\includegraphics[width=5.65cm]{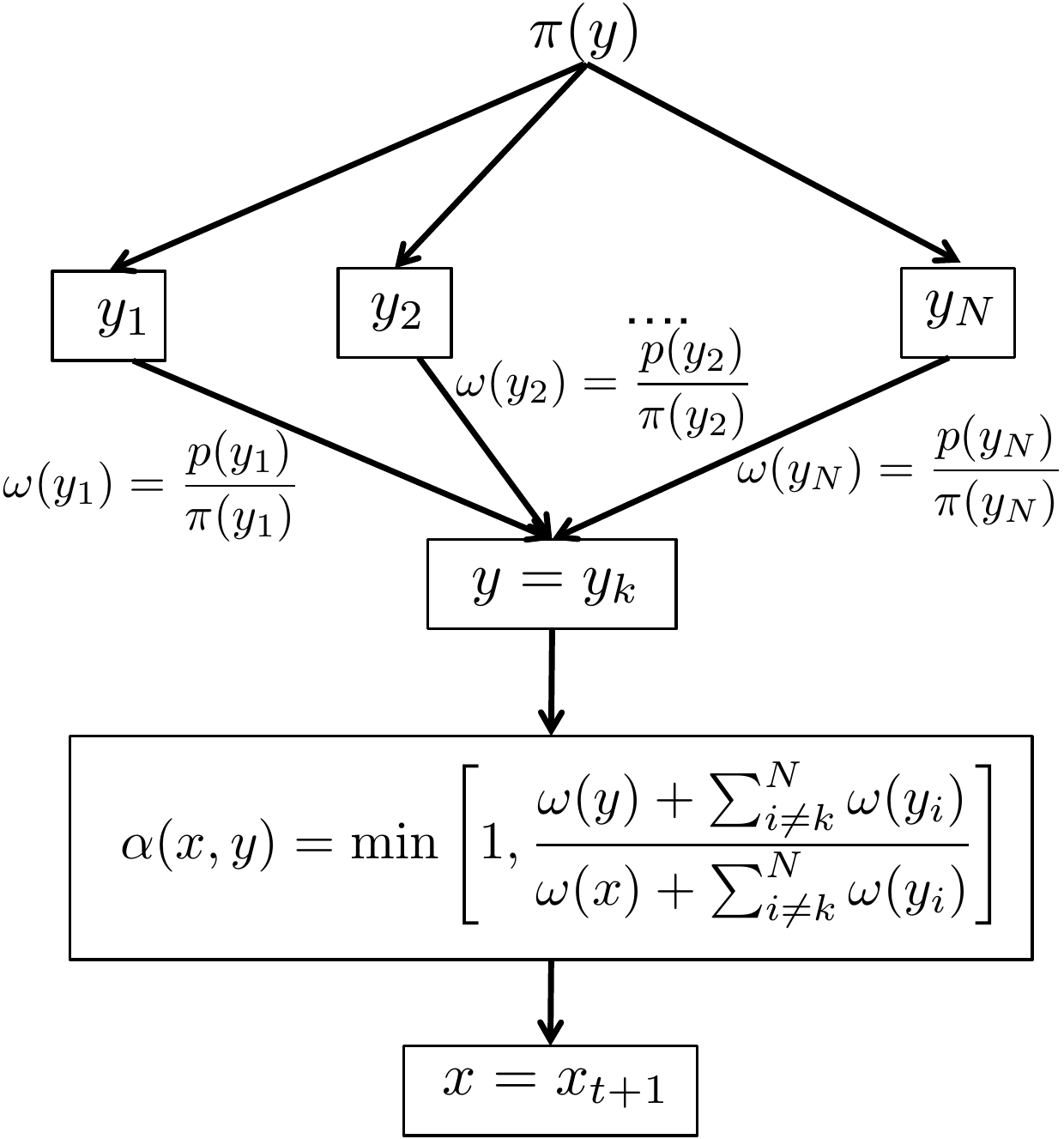}}
		\subfigure[]{\includegraphics[width=5.7cm]{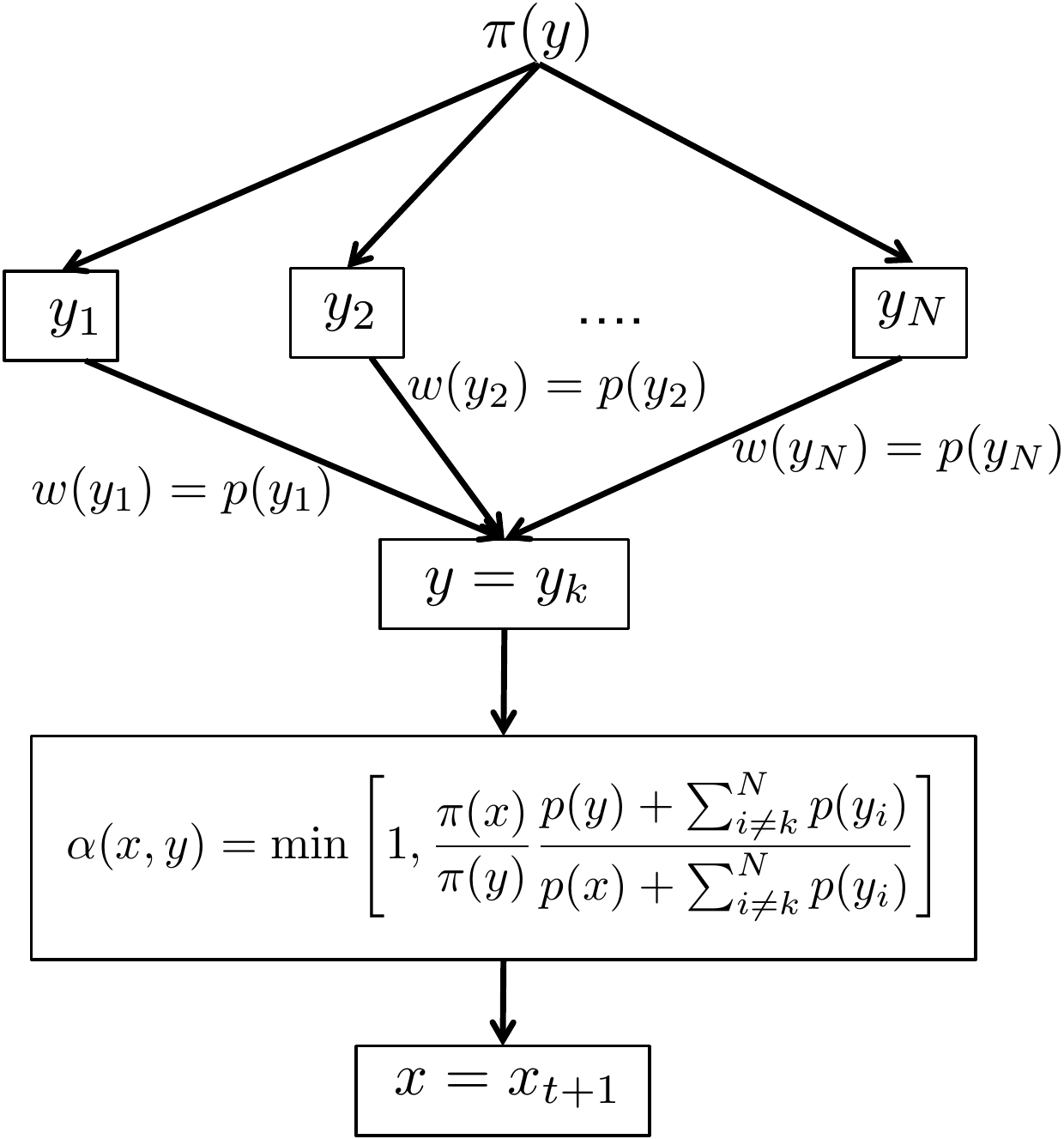}}		 
 		}
\caption{Sketch of the simplest MTM schemes using just one independent proposal density, {\bf (a)} with importance weights and {\bf (b)} weights proportional to $p(x)$. In these cases, clearly $\alpha(x,y)\rightarrow 1$ as $N\rightarrow \infty$.}
\label{fig4}
\end{figure}

Figure \ref{fig3} depicts the scheme of a MTM with generic weights and different independent proposal pdfs, whereas Figure \ref{fig4} shows virtually the simplest MTM algorithms, using the same independent proposal to draw the $N$ candidates and importance weights (Fig. \ref{fig4}(a)) or weights proportional to the target (Fig. \ref{fig4}(b)).\footnote{Another simple MTM scheme is the ``orientational bias Monte Carlo'' \citep[Chapter 13]{Frenkel96}. In this case, the proposal pdf must be symmetric, i.e., $\pi(y|x)=\pi(x|y)$, and the weights must be proportional to the target, i.e., $\omega(y_i)=p(y_i)$, $i=1,...,N$. } In this special cases, the analysis of the algorithm is also simpler. Indeed, for instance, consider the case in Fig. \ref{fig4}(a). The acceptance probability can be expressed as 
$$\alpha(x,y)=\min\left[1,\frac{\omega(y)+\sum_{i\neq k}^{N}\omega(y_i)}{\omega(x)+\sum_{i\neq k}^{N}\omega(y_i)}\right],$$ 
where $w(y_i)=\frac{p(y_i)}{\pi(y_i)}.$
Note that, in this case clearly $\alpha(x,y)\rightarrow 1$ as $N\rightarrow \infty$, since the chosen candidate is ``extremely good'' using the importance sampling principle, when $N\rightarrow \infty$. 

%%%%%%%%%%%%%%%%
%%%%%%%%%%%%%%%%
\section{Numerical simulations}
%%%%%%%%%%%%%%%%
%%%%%%%%%%%%%%%%
\label{ToyExample}

In this section, we provide numerical results comparing different MTM approaches: using random walks or independent proposal pdfs, with different weight functions, without drawing the reference points and using different acceptance functions. All the results have been averaged over $2000$ runs and they are obtained generating $5000$ iterations of the Markov chain, with the exception of the last example where we only draw $500$ samples.

%%%%%%%%%%%%%%%%%%%%%%%%
\subsection{Random walk proposal densities}
%%%%%%%%%%%%%%%%%%%%%%%%

Let $X\in \mathbb{R}$ be a random variable\footnote{Note that, in this work, we have mainly considered scalar variables in order to simplify the treatment and the notation. All the considerations and algorithms contained in this work are also valid for multi-dimensional variables (see, for instance, the last numerical example in Section \ref{SmilingFace}).} with bimodal pdf 
\begin{equation}
\label{targetToy}
p_o(x)\propto p(x)= \exp\left\{-(x^2-4)^2/4 \right\}=\exp\left\{-\frac{x^4-8x^2+16}{4} \right\}.
\end{equation}
We want to draw samples from $p_o(x)$ using different MTM schemes. We generate tries from a Gaussian proposal with variance $\sigma^2$ and the mean depends on the previous state $x$ of the chain, i.e.,
\begin{equation}
\label{RWGaussian}
\pi(y|x) \propto \exp\left\{-\frac{(y-x)^2}{2\sigma^2}\right\}.
\end{equation}
We apply MTM methods using the proposal above, different number of candidates $N=1,2,5,100,1000$ and different standard deviation $\sigma=2,10$. Importance weights $\omega(y_i,x)=\frac{p(y_i)}{\pi(y_i|x)}$ are used to select a good candidate.
Observe that an MTM with $N=1$ is exactly a standard MH algorithm. We also apply different MTM techniques without drawing the reference points (denoted as ``MTM-without'') described in Section \ref{MTMwithoutReferencepoints}. Tables \ref{TablaResults1} and \ref{TablaResults2} summarize the numerical results in terms of averaged probability of accepting a movement and linear correlation between the state $x_{t}$ and $x_{t+1}$.
\begin{table}[!hbt]
\begin{center}
\caption{Numerical results (proposal as random walk, $\sigma=2$, using importance weights).}
\label{TablaResults1}
\begin{tabular}{|c|c|c|c|} 
\hline
{\bf Technique} & {\bf Number of tries} & {\bf Acceptance rate} & {\bf Linear correlation}  \\ 
\hline
\hline
standard MH  & $N=1$ &  0.3002 &   0.9053 \\ 
(MTM with $N=1$) & & & \\
\hline
MTM-rw  & $N=2$ &  0.4363 &  0.8397  \\ 
\hline
MTM-rw & $N=5$ & 0.6046 &  0.6989  \\ 
\hline
MTM-rw &$N=100$  & 0.8647 &  0.1892  \\ 
\hline
MTM-rw &$N=1000$  & 0.9557 & 0.0513 \\ 
\hline
\hline
MTM-without &$N=2$  & 0.4229 & 0.9160 \\ 
\hline
MTM-without &$N=5$  & 0.5121 & 0.9568 \\
\hline
MTM-without &$N=100$  & 0.1902 & 0.9978 \\
\hline
MTM-without &$N=1000$  & 0.0036 & 0.9993 \\
\hline
\end{tabular}
\end{center}
\end{table}

\begin{table}[!hbt]
\begin{center}
\caption{Numerical results (proposal as random walk, $\sigma=10$, using importance weights).}
\label{TablaResults2}
\begin{tabular}{|c|c|c|c|} 
\hline
{\bf Technique} & {\bf Number of tries} & {\bf Acceptance rate} & {\bf Linear correlation}  \\ 
\hline
\hline
standard MH  & $N=1$ &  0.0991 &  0.9085 \\ 
(MTM with $N=1$) & & & \\
\hline
MTM-rw  & $N=2$ &   0.1795 &  0.8335  \\ 
\hline
MTM-rw & $N=5$ & 0.3483 & 0.6700  \\ 
\hline
MTM-rw &$N=100$  & 0.8373 & 0.1676 \\ 
\hline
MTM-rw &$N=1000$  & 0.9483 & 0.0522  \\ 
\hline
\hline
MTM-without &$N=2$  & 0.1810 & 0.8376 \\ 
\hline
MTM-without &$N=5$  & 0.3575 & 0.7017 \\
\hline
MTM-without &$N=100$  &  0.4453 & 0.9264 \\
\hline
MTM-without &$N=1000$  & 0.2612 &  0.9952 \\
\hline
\end{tabular}
\end{center}
\end{table}

 It is important to remark that high acceptance rates are not a suitable indicator of good performance since, in general, the best acceptance rate is different from $1$ \citep{Roberts97}. Therefore, better performance is indicated by smaller correlations. We show also the acceptance rates because of the MTM method (drawing the reference points) presents a behavior typical in adaptive MCMC algorithms where the adaptive proposal pdf convergence to the true shape of the target \citep{MartinoA2RMS}: the acceptance rate grows and the linear correlation decreases quickly as $N\rightarrow +\infty$.
 Indeed, we can observe that, in both cases $\sigma=2,10$, the correlation obtained with the MTM decreases to zero as $N\rightarrow +\infty$. 
 Without drawing the reference points, the resulting algorithm is totally useless for $\sigma=2$ (Table \ref{TablaResults1}) whereas it outperforms the standard MH for $N=2$ and $N=5$ for $\sigma=10$ (Table \ref{TablaResults2}). However, increasing $N$ the performance gets worse. The results in Table \ref{TablaResults2} suggest that it exists an {\it optimal} number of tries for an MTM scheme without generating randomly the reference points. However, the MTM method with the additional cost of  the random generation of reference points always outperforms the general scheme described in Section \ref{MTMwithoutReferencepoints}. With independent proposal pdfs this is not true as we show later. 
    
%with another target pdf

%%%%%%%%%%%%%%%%%%%%%%%
\subsection{Different choice of the weights}
%%%%%%%%%%%%%%%%%%%%%%%
\label{DiffWeightsSect}
Considering the same target pdf in Eq. \eqref{targetToy}, the Gaussian proposal with $\sigma=10$ in Eq. \eqref{RWGaussian} (random walk) and using $N=100$ tries,
we have compared the performance of different weight functions. Table \ref{TablaResults3} summarizes the results. 
\begin{table}[!hbt]
\begin{center}
\caption{Numerical results (proposal as random walk, $\sigma=10$, $N=100$ tries).}
\label{TablaResults3}
\begin{tabular}{|c|c|c|} 
\hline
{\bf Weights} & {\bf Acceptance rate} & {\bf Linear correlation}  \\ 
\hline
\hline
$\omega_i(y_i,x)=\frac{p(y_i)}{\pi_i(y_i|x)}$  &   0.8373 & 0.1676  \\ 
importance weights &   &    \\ 
\hline
$\omega_i(y_i,x)=p(y_i)$ & 0.8374 & 0.1959   \\ 
\hline
$\omega_i(y_i,x)=1$ & 0.0988 & 0.9090  \\ 
\hline
$\omega_i(y_i,x)=\sqrt{p(y_i)}$&  0.7036 &  0.3340\\ 
\hline
$\omega_i(y_i,x)=[p(y_i)]^2$ & 0.6870 &  0.3093  \\ 
\hline
$\omega_i(y_i,x)=[p(y_i)]^3$ & 0.4476 & 0.4020  \\ 
\hline
$\omega_i(y_i,x)=\pi_i(x|y_i)$ & 0.1348  &  0.8809 \\ 
\hline
$\omega_i(y_i,x)=\frac{1}{\pi_i(y_i|x)}$ & 0.0365  &  0.9652 \\ 
\hline
$\omega_i(y_i,x)=p(y_i)\pi_i(x|y_i)$ &  0.8371  & 0.2248  \\ 
\hline
\end{tabular}
\end{center}
\end{table}

The best results are provided by the importance weights $\omega_i(y_i,x)=\frac{p(y_i)}{\pi_i(y_i|x)}$. The weights of the form $\omega_i(y_i,x)=p(y_i)$  and $\omega_i(y_i,x)=p(y_i)\pi_i(x|y_i)$ also yield small correlation. Clearly,  the choice $\omega_i(y_i,x)=1$ produces the same results of a standard MH since the selected candidate is chosen uniformly among the set of drawn tries $y_i$, $i=1,...,N$, without using any information of the target or the proposal functions.

%%%%%%%%%%%%%%%%%%%%%%%
\subsection{Independent proposal densities}
%%%%%%%%%%%%%%%%%%%%%%%
\label{IDPROP}
In order to draw samples from the target in Eq. \eqref{targetToy}, we also apply MTM algorithms with independent proposal  densities (MTM-ind) as
\begin{equation}
%\label{INGaussian}
\nonumber
\pi(y) \propto \exp\left\{-\frac{(y-\mu)^2}{2\sigma^2}\right\},
\end{equation}
with $\sigma=10$. In a first scheme, we generate $N=100$ candidates from one proposal with $\mu=0$. Moreover, in other scheme, we use two different independent proposal pdfs with  
$\mu_1=-10$ and $\mu_2=2$. In this case, we draw $N/2=50$ tries from each one. We apply these schemes with importance weights, $\omega_i(y_i,x)=\frac{p(y_i)}{\pi_i(y_i)}$, and also with weights just proportional to the target pdf, $\omega_i(y_i,x)=p(y_i)$. Table \ref{TablaResults4} shows the numerical results.
\begin{table}[!hbt]
\begin{center}
\caption{Numerical results ($\sigma=10$, $N=100$ tries).}
\label{TablaResults4}
\begin{tabular}{|c|c|c|} 
\hline
{\bf Proposal pdfs} & {\bf Acceptance rate} & {\bf Linear correlation}  \\ 
\hline
\hline
MTM-rw  with &   0.8373 & 0.1676  \\ 
$\omega_i(y_i,x)=\frac{p(y_i)}{\pi_i(y_i|x)}$&   &    \\ 
\hline
MTM-rw  with &   0.8374 & 0.1959  \\ 
$\omega_i(y_i,x)=p(y_i)$&   &    \\ 
\hline
MTM-ind  with &   0.9760 & 0.0252  \\ 
one proposal pdf ($\mu=0$) and &    &   \\ 
$\omega_i(y_i,x)=\frac{p(y_i)}{\pi_i(y_i|x)}$ &   &    \\ 
\hline
MTM-ind  with &   0.9751 & 0.0267  \\ 
one proposal pdf ($\mu=0$) and &    &   \\ 
$\omega_i(y_i,x)=p(y_i)$ &   &    \\ 
\hline
MTM-ind  with &   0.7420 & 0.2748 \\ 
two proposal pdfs ($\mu_1=-10$ and $\mu_2=2$) &    &   \\ 
and $\omega_i(y_i,x)=\frac{p(y_i)}{\pi_i(y_i|x)}$ &   &    \\ 
\hline
MTM-ind  with &   0.7509 &  0.6622  \\ 
two proposal pdfs ($\mu_1=-10$ and $\mu_2=2$) &    &   \\ 
and $\omega_i(y_i,x)=p(y_i)$ &   &    \\ 
\hline
\end{tabular}
\end{center}
\end{table}

The first two lines of the Table \ref{TablaResults4} recall the acceptance rates and the linear correlations using the random walk proposal densities.
The table shows that the MTM with independent proposal with $\mu=0$ provides the best results, i.e., the smallest correlation. However, the results depend strongly on a suitable tuning of the parameter $\mu$. Also in this case, the importance weights seem to provide better results. 
Another important consideration is that, using two proposal pdfs, the MTM has selected a candidate generated from the proposal with $\mu_1=-10$ with a rate of   $39.5 \%$ using importance weights, and just $1.5 \%$ with the weights proportional to the target. This observation can be extremely important to design an adaptive strategy where the best proposal density is chosen among of a set of proposals.

%%%%%%%%%%%%%
\subsection{Heavy tails}
%%%%%%%%%%%%%
\label{HeavyTail}

In order to analyze the performance of the MTM schemes with heavy tails, now we consider as target pdf  the so-called  {\em L\'evy distribution} for non-negative random variable, namely,
\begin{equation}
\label{LevyTarget}
p_o(x)\propto p(x)=\frac{1}{(x-\eta)^{3/2}}\exp\left(-\frac{\nu}{2(x-\eta)}\right), \mbox{ }  \forall x\geq \eta\geq 0.
\end{equation}
The normalizing constant $\frac{1}{c_p}$, such that $p_o(x) = \frac{1}{c_p} p(x)$, is analytically known, $\frac{1}{c_p} = \sqrt{\frac{\nu}{2\pi}}$.
Moreover, given a random variable $X\sim p_o(x)$, all the moments $E[X^{\gamma}]$ with $\gamma\geq 1$ do not exist owing to the heavy tail characteristic of the L\'evy distribution.

Our goal is to estimate the normalizing constant $\frac{1}{c_p}$ via Monte Carlo simulation, when $\eta=0$ and $\nu=2$, generating $5000$ iterations of the Markov chain.
We apply three different MTM techniques with $N=1000$ tries (drawing the reference points) and using importance weights to choose a suitable candidate each step. In the first two schemes (MTM-ind), we use an independent proposal $\pi(x_t)\propto \exp\{-(x_t-\mu)^2/(2\sigma^2)\}$ with $\mu=10, 100$ and $\sigma=50$, whereas, in the last one (MTM-rw), we use a random walk proposal $\pi(x_t|x_{t-1})\propto \exp\{-(x_t-x_{t-1})^2/(2\sigma^2)\}$ with  $\sigma=50$. We choose huge values of  $\sigma$ due to the heavy tail feature of the target.
We have averaged all the results over $2000$ runs and they are summarized in Table \ref{resultsotherEjootro}. The real value of $\frac{1}{c_p}$ when $\nu=2$ is  $\sqrt{\frac{2}{2\pi}}=0.5642$.\footnote{We do not provide the estimated linear correlation because of the moments (as the mean, for instance) of the target do not exist, and it makes difficult a right estimation of the correlation.}
%our estimation is $0.5652$ with std $0.0014$ and real value is $\sqrt{\frac{2}{2\pi}}=0.5642$.
\begin{table*}[!hbt]
\def\marginwidth{1.5mm}
\begin{center}
\caption{Estimation of the constant $\frac{1}{c_p} = \sqrt{\frac{2}{2\pi}}=0.5642$ and standard deviation of the estimation ($N=1000$ tries).}
\label{resultsotherEjootro}
\begin{tabular}{|c@{\hspace{\marginwidth}}|c@{\hspace{\marginwidth}}|c@{\hspace{\marginwidth}}|c@{\hspace{\marginwidth}}|}
\hline
{\bf Technique}  &{\bf Estimation} &{\bf Std of} & {\bf Further}  {\bf informations}\\
  &{\bf of $\frac{1}{c_p}$} &{\bf the estimation}  &\\
\hline
\hline
MTM-ind & 0.6056 & 0.0012 &  $\mu=10$, $\sigma= 50$ \\
\hline
MTM-ind & 0.5994 & 0.0010 & $\mu=100$, $\sigma= 50$ \\
\hline
MTM-rw & 0.5819 & 0.0050 & $\sigma=50$\\
\hline
\end{tabular}
\end{center}
\end{table*}

%%%%%%%%%%%%%%%%%%%%%%%%%
\subsection{Different acceptance probabilities}
%%%%%%%%%%%%%%%%%%%%%%%%%
%We consider a simple example to provide a ``numerical proof''. We recall that the theoretical proof is given in Section \ref{SectProof}.
%Note that in this
In this section, we consider again the bimodal target density in Eq. \eqref{targetToy}, i.e.,  $p_o(x)\propto p(x)= \exp\left\{-(x^2-4)^2/4 \right\}$, and we generate candidates from a random walk Gaussian density with $\sigma=1$, i.e., $\pi(y|x) \propto \exp\left\{-\frac{(y-x)^2}{2}\right\}$.   
We choose as weight functions $\omega(x,y)=[p(x)]^{\theta}$, with $\theta=1/2$. Note that they cannot be obtained using the analytic form necessary in the standard MTM \citep{Liu00}.
Moreover, we consider four possible combinations of the $\beta(x,y)$ and $\gamma(x,y)$ functions
\begin{gather}
\nonumber
\begin{split}
&\alpha_{1,1}(x,y)=\beta_1(x,y)\gamma_1(x,y), \\
& \alpha_{1,2}(x,y)=\beta_1(x,y)\gamma_2(x,y), \\
&\alpha_{1,3}(x,y)=\beta_1(x,y)\gamma_3(x,y), \\
&\alpha_{2,3}(x,y)=\beta_2(x,y)\gamma_3(x,y), \\
\end{split}
\end{gather}
where each $\beta_i(x,y)$,  $i=1,2$, and $\gamma_j(x,y)$,  $j=1,2,3$, are defined in Sections \ref{SectBeta} and \ref{SectGamma}.  
Then, we run the different MTM algorithms with $N=10$ and $N=100$ candidates.
%Figure \ref{fig1} illustrates the target density $p_o(x)$ (solid line) and the normalized histogram of $100,000$ samples drawn from the MTM scheme using $\alpha_{1,3}(x,y)$ and $N=10$. We can observe that the histogram approximates closely the shape of the target pdf, i.e., the Markov chain generated by the novel scheme converges to $p_o(x)$. 
%\begin{figure}[htb]
%\centerline{
% 		\includegraphics[width=4.5cm]{Fig1.pdf} 
% 		}
%\caption{The target density $p_o(x) \propto \exp\left\{-(x^2-4)^2/4 \right\}$ (solid line) and the normalized histogram of $100,000$ samples drawn from the extended MTM technique using $\alpha_{1,3}(x,y)$, $N=10$ and weights $\omega(x,y)=\sqrt{p(x)}$.}
%\label{fig1}
%\end{figure} 
 Table \ref{TablaResults} shows the acceptance rate (the averaged probability of accepting a movement) and normalized linear correlation coefficient (between one state of the chain and the next) averaged over $2000$ runs and obtained with the different techniques where $N=10$. 
\begin{table}[!hbt]
\begin{center}
\caption{Numerical results with $N=10$.}
\label{TablaResults}
\begin{tabular}{|c|c|c|} 
\hline
{\bf Function $\alpha$} & {\bf Acceptance rate} & {\bf Linear correlation}  \\ 
\hline
\hline
$\alpha_{1,1}(x,y)$ &  0.1167 &  0.9932 \\ 
\hline
$\alpha_{1,2}(x,y)$ &  0.3246 &  0.9811  \\ 
\hline
$\alpha_{1,3}(x,y)$ & 0.5512 & 0.9756  \\ 
\hline
$\alpha_{2,3}(x,y)$ & 0.3370 & 0.9806  \\ 
\hline
\end{tabular}
\end{center}
\end{table}

Table \ref{TablaResults22} illustrates the results using $N=100$.  We observe that $\alpha_{1,3}$ provides that  greatest acceptance rate and lowest correlation in both cases.  The acceptance rate of $\alpha_{1,1}$ decreases with $N=100$ because of $\gamma_1(x,y|{\bf x}^{*}_{-k},{\bf y}_{-k})=W_x$ diminishes with the number of tries $N$.
Moreover, the correlation appears (almost) invariant with the number of tries $N$.
\begin{table}[!hbt]
\begin{center}
\caption{Numerical results with $N=100$.}
\label{TablaResults22}
\begin{tabular}{|c|c|c|} 
\hline
{\bf Function $\alpha$} & {\bf Acceptance rate} & {\bf Linear correlation}   \\
\hline
\hline
$\alpha_{1,1}(x,y)$ &  0.0173 &  0.9931 \\ 
\hline
$\alpha_{1,2}(x,y)$ &  0.3354 &  0.9828  \\ 
\hline
$\alpha_{1,3}(x,y)$ & 0.5904 & 0.9737  \\ 
\hline
$\alpha_{2,3}(x,y)$ & 0.3540 & 0.9859  \\ 
\hline
\end{tabular}
\end{center}
\end{table}

Better performances can be attained using the acceptance function of \citep{Pandolfi10} and rewritten in Eq. (\ref{alpha1trad}), as expected analyzing the analytic form of the different  acceptance functions. Indeed, we obtain acceptance rates of $0.74$, $0.81$ and correlation $0.96$, $0.96$ with $N=10$ and $N=100$, respectively.  
%However, observe again that high acceptance rates for MCMC techniques are not in general desirable \citep{Roberts97}.

% Indeed, performances is also affected by the choice of the proposal pdf (not just for shape but also for the type of dependence with the previous state) and by the specific shape of the target functions, for instance. 

%%%%%%%%%%%%%%%%%%%%%%%%%
\subsection{Smiling-Face distribution}
\begin{figure}[htb]
\centerline{
 		\subfigure[]{\includegraphics[width=6cm]{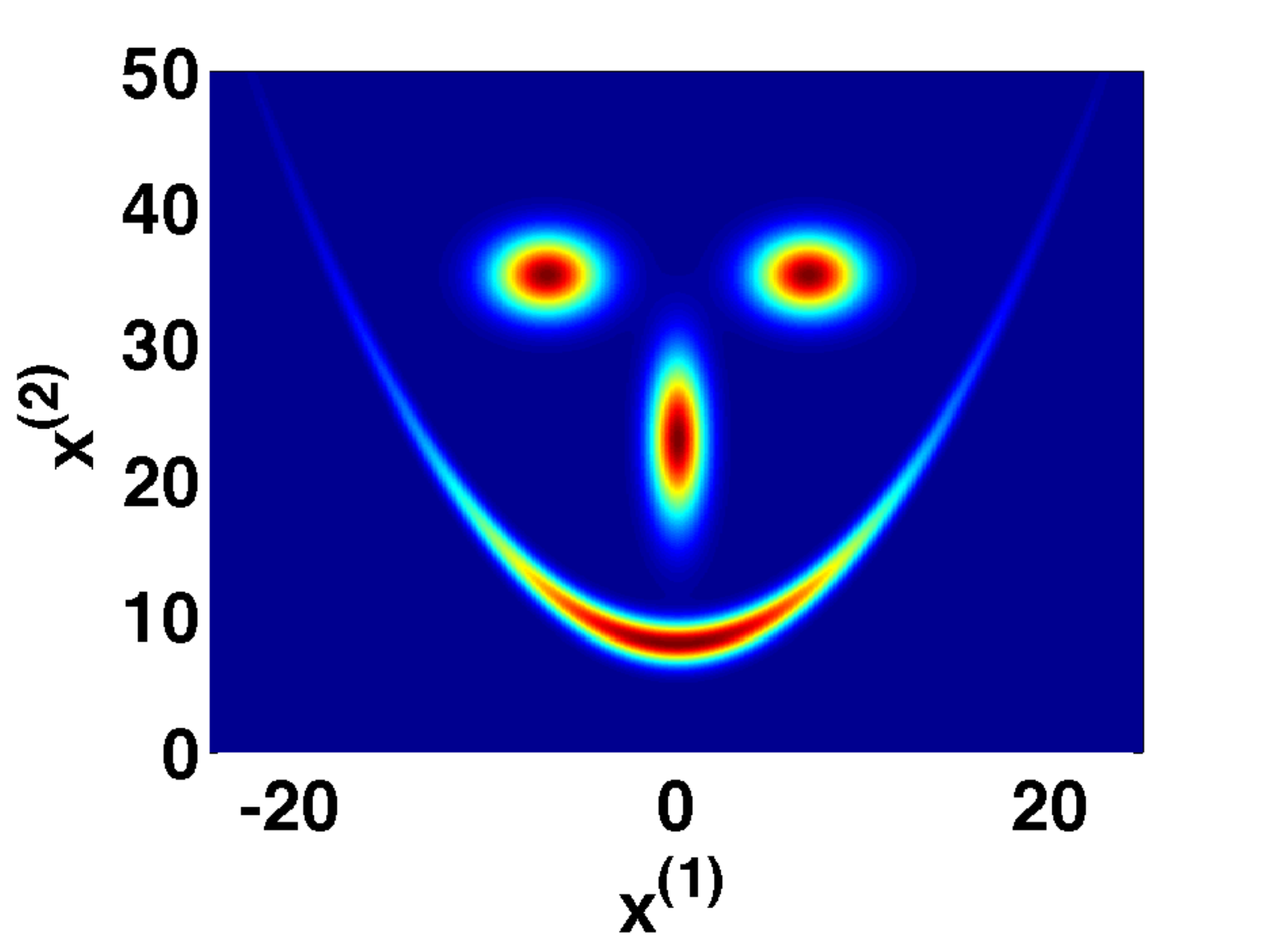}} 
		}
\centerline{
		\subfigure[]{\includegraphics[width=6cm]{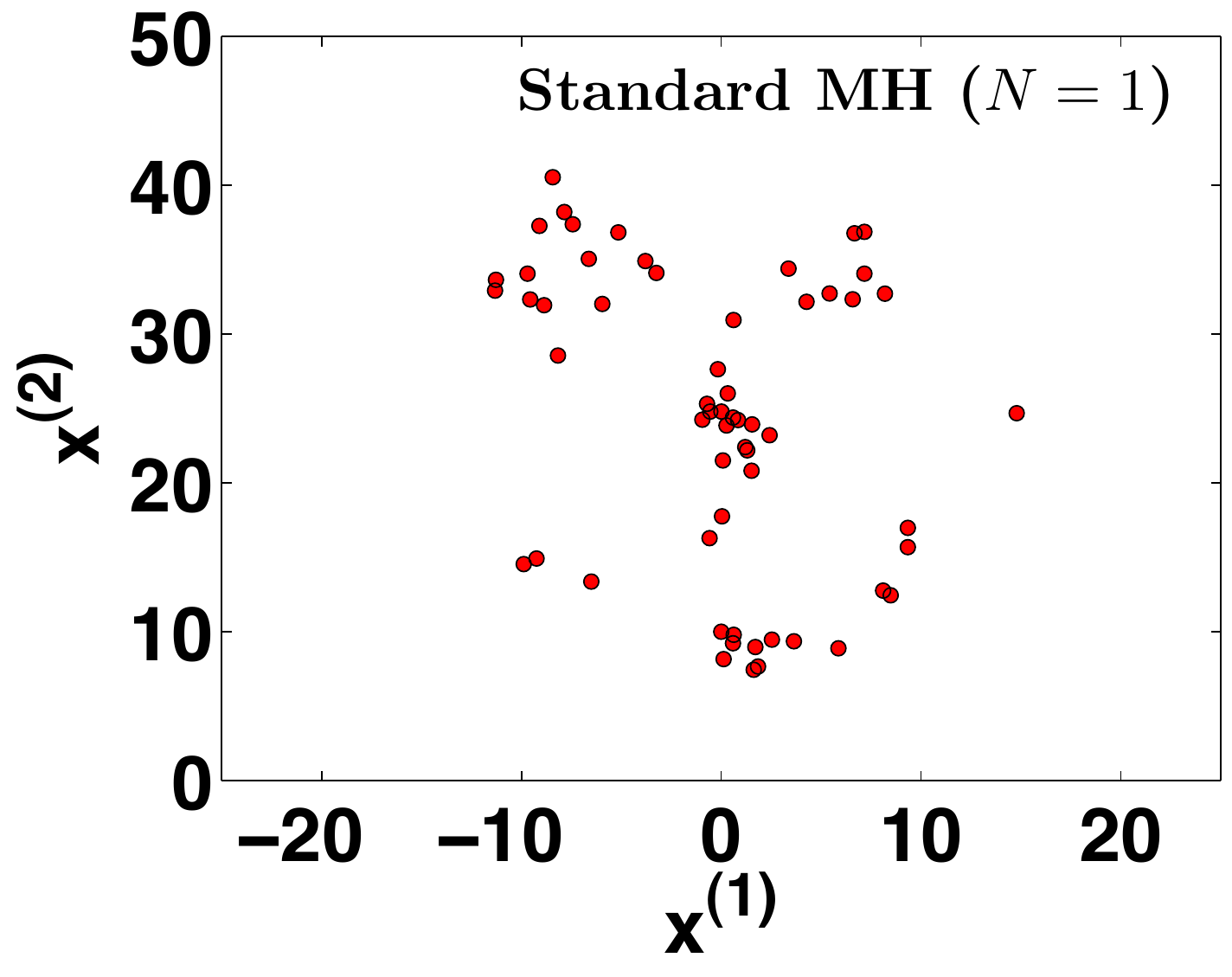}}
			\subfigure[]{\includegraphics[width=6cm]{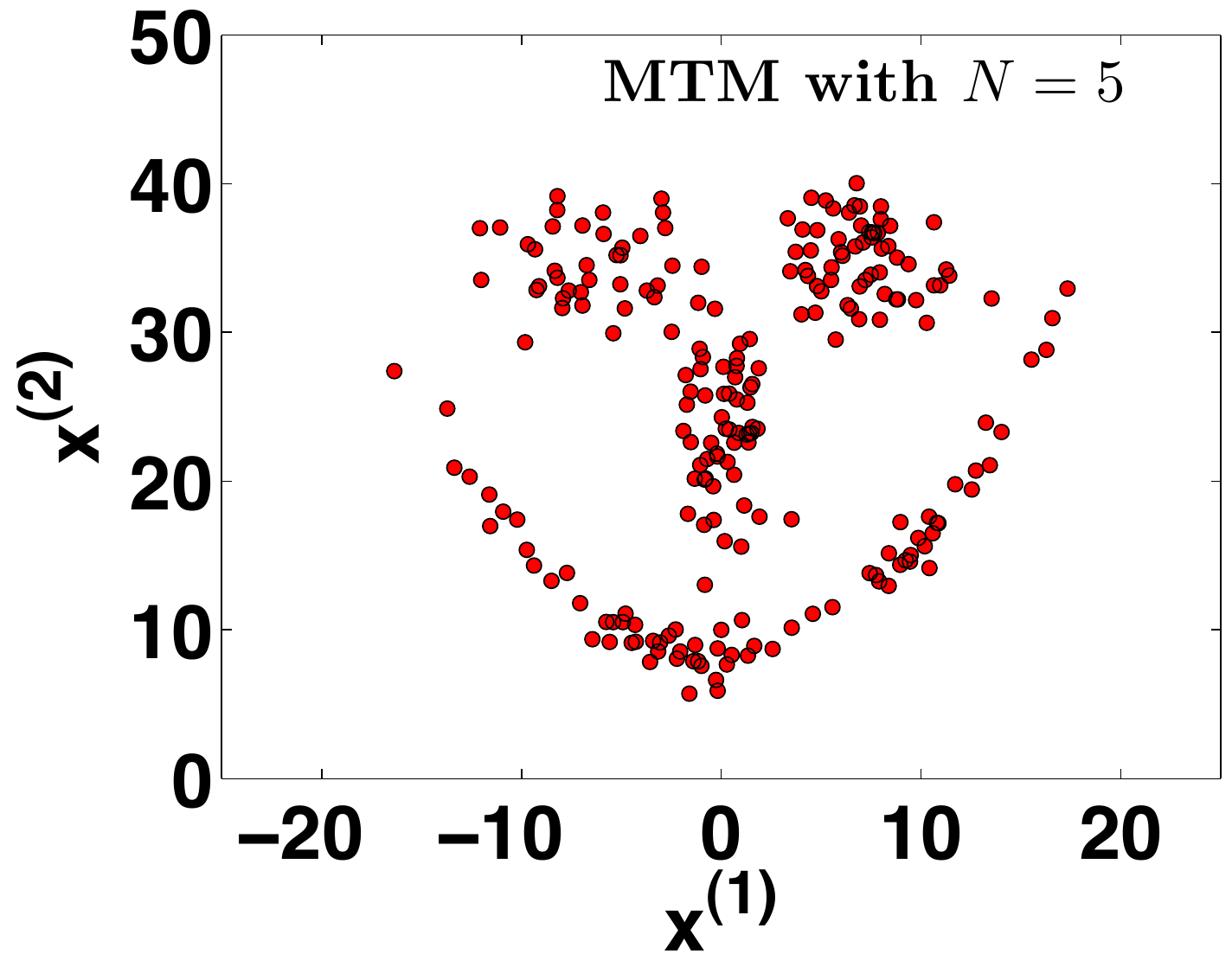}}
 	 		}
\centerline{
		\subfigure[]{\includegraphics[width=6cm]{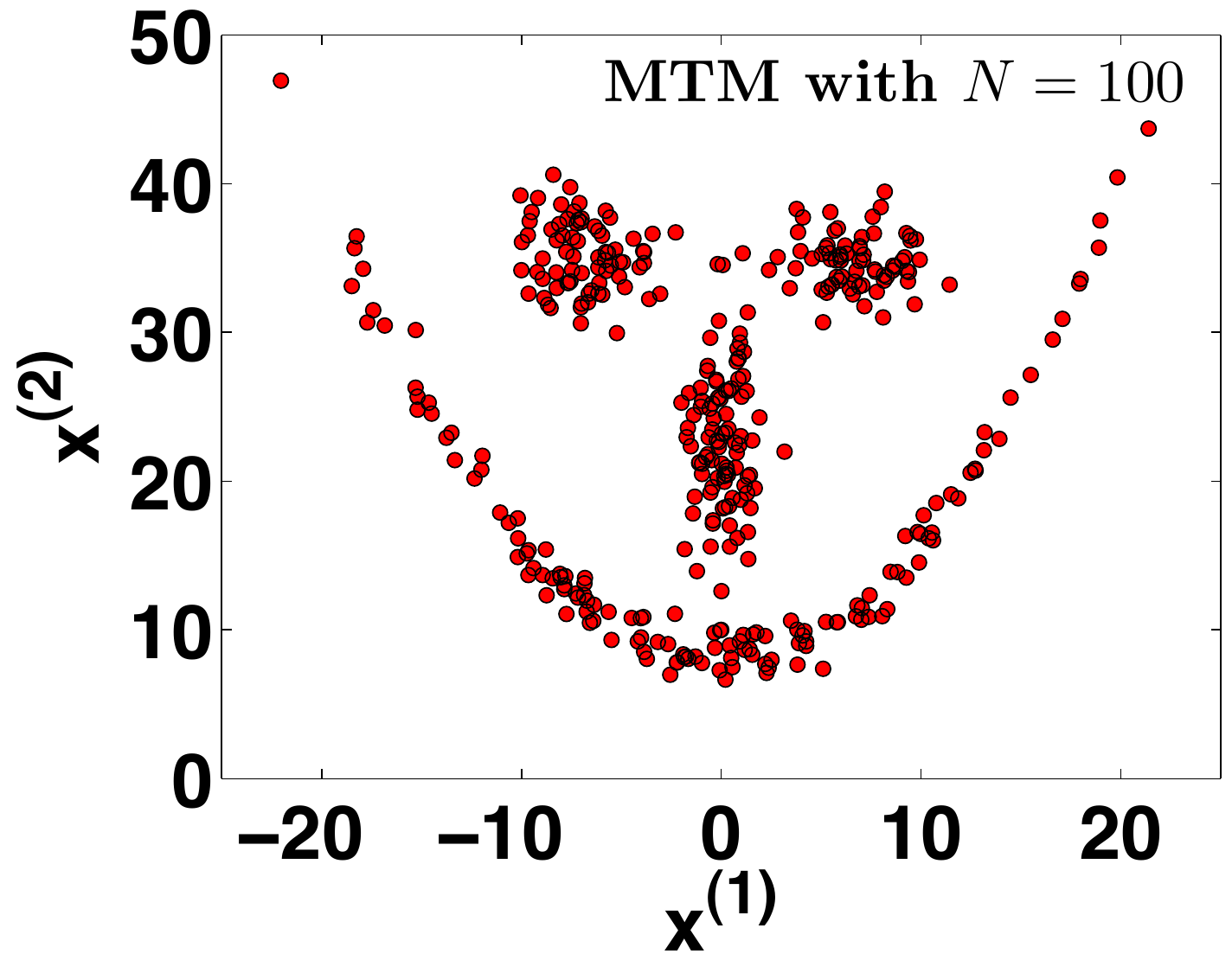}}
	       \subfigure[]{\includegraphics[width=6cm]{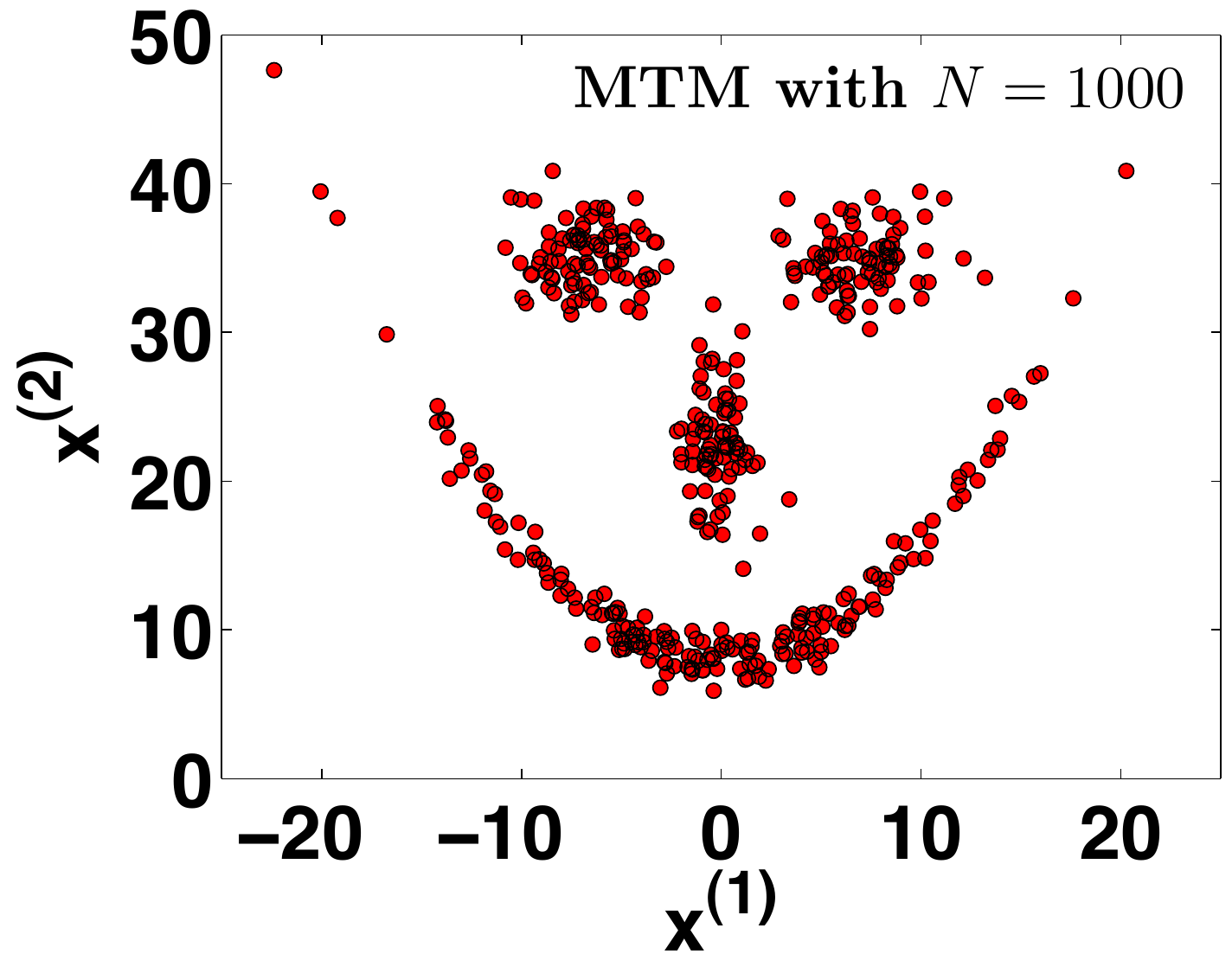}}
 	 		}
			
\caption{ {\bf (a)} The Smiling-Face target density. The remaining figures {\bf (b)}-{\bf (c)}-{\bf (d)}-{\bf (e)} depict the first $500$ generated samples drawn from the different samplers in one run (with $\sigma_p=10$). Note that the number of points are less than $500$ since, in certain iterations, the chain remains in the same state (depending on the acceptance probability $\alpha$) so that some points are repeated. {\bf (b)} Samples generated by a standard MH ($N=1$). {\bf (c)} Samples generated by a MTM with $N=5$. {\bf (d)} Samples generated by a MTM with $N=100$. {\bf (e)} Samples generated by a MTM with $N=1000$. It is evident that the MTM scheme speeds up the convergence of the Markov chain.}
\label{figLastIhope}
\end{figure}

%%%%%%%%%%%%%%%%%%%%%%%%%
\label{SmilingFace}
In this section, we show that the power of the MTM schemes increases when they draw from more complicated target distributions in higher dimensions, w.r.t.\ a standard MH algorithm. To provide a graphical example, we consider a bidimensional target pdf $p_o({\bf x})$ (where ${\bf x}=[x^{(1)},x^{(2)}]^T\in \mathbb{R}^2$, $x^{(i)}\in\mathbb{R}$, $i=1,2$) composed as a mixture of $4$ densities, 
\begin{equation}
\label{TargetSmile}
p_o({\bf x})\propto \frac{1}{4}\sum_{i=1}^{4}  p_i({\bf x}). 
\end{equation}
The first three components are proportional to bivariate Gaussian pdfs, i.e.,
$$p_i({\bf x})=p_i(x^{(1)},x^{(2)})= \exp\left\{-\frac{\left(x^{(1)}-\mu_i^{(1)}\right)^2}{2\left(\sigma_i^{(1)}\right)^2}-\frac{\left(x^{(2)}-\mu_i^{(2)}\right)^2}{2\left(\sigma_i^{(2)}\right)^2} \right\},$$ 
with $i=1,2,3$, $\mu_1^{(1)}=-7$,  $\mu_1^{(2)}=35$, $\mu_2^{(1)}=7$,  $\mu_2^{(2)}=35$, $\mu_3^{(1)}=0$ , $\mu_3^{(2)}=23$, $\sigma_1^{(1)}=2$, $\sigma_1^{(2)}=2$, $\sigma_2^{(1)}=2$, $\sigma_2^{(2)}=2$,  $\sigma_3^{(1)}=1$ and  $\sigma_3^{(2)}=4$. 
The last component is a banana-shaped density \citep{Haario99,Lan12}, i.e., 
$$
p_4({\bf x})=p_4(x^{(1)},x^{(2)})= \exp\left\{-\frac{\left(x^{(1)}\right)^2}{\eta}-\frac{\left(x^{(1)}-\rho\left(x^{(2)}\right)^2+100\rho\right)^2}{2} \right\},
$$
with $\eta=144.5$ and $\rho=0.08$. The banana-shaped distribution was first introduced in \citep{Haario99} and is known in literature to be a difficult target.  
This kind of bidimensional and multimodal mixtures of densities is often used to compare the performance of different MCMC techniques \citep[Chapter 5]{Liang10}, \citep{Haario99,Haario01,Lan12}. %Numerical examples with bidimensional pdfs ...are often used....with this kind of target.... 
The parameters of the Gaussian components and the banana-shaped pdf are chosen in order to form a ``smiling face'' as illustrated in Figure \ref{figLastIhope}(a). The reason is that, in this way, it is possible to illustrate {\it graphically} the performance of different samplers, as we show below.  
%\newpage

To draw from $p_o({\bf x})$, we apply a MH and a MTM scheme using for both a random walk Gaussian proposal pdf, i.e.,
$$\pi({\bf x}_t|{\bf x}_{t-1})\propto \exp \left\{-\left(x^{(1)}_t-x^{(1)}_{t-1}\right)^2/(2\sigma_p^2)-\left(x^{(2)}_t-x^{(2)}_{t-1}\right)^2/(2\sigma_p^2)\right\}.$$ 
In order to show the speed of the convergence of the samplers, we have generated only $500$ samples with a MTM with different number of candidates $N=1,5,100,1000$ (note with $N=1$ is a standard MH) and different standard deviation $\sigma_p=5,10$ of the proposal.
 
Tables \ref{TablaLast1}-\ref{TablaLast2} provide the average acceptance probability of a new state in the first column (the averaged values of $\alpha$),  the jump rate among different modes in the second column (from ``left eye'' to the ``smile'', or from the ``smile'' to the ``nose'' etc.) and the linear correlation for each component of ${\bf x}$, in the last column.     
To compute the mode-jump rate we establish that the state ${\bf x}_t$ belongs to the mode $i^*$ if
$$
i^*=\arg \max_{i\in \{1,...,4\}} p_i({\bf x}_t),
$$
where $p_i({\bf x}_t)$ are the $4$ components in the mixture of Eq. \eqref{TargetSmile}.
All results are averaged over $2000$ runs using $\sigma_p=5$ in Table \ref{TablaLast1} and $\sigma_p=10$ in Table \ref{TablaLast2}.
\begin{table}[!hbt]
\begin{center}
\caption{Numerical results with $\sigma_p=5$.}
\label{TablaLast1}
\begin{tabular}{|c|c|c|c|} 
\hline
{\bf Number of tries $N$} & {\bf Acceptance Rate} & {\bf Mode-Jump Rate} & {\bf Correlation}   \\
\hline
\hline
$N=1$ (standard MH) &   0.2296 &  0.0401 & $x^{(1)}\rightarrow$ 0.9460 \\ 
 &    &   & $x^{(2)}\rightarrow$ 0.9749 \\ 
\hline
$N=5$ &  0.5118 &  0.1166 &  $x^{(1)}\rightarrow$  0.8661 \\ 
 &    &   & $x^{(2)}\rightarrow$ 0.9492 \\ 
\hline
$N=100$ & 0.7137 & 0.3373  &  $x^{(1)}\rightarrow$ 0.6193 \\
 &    &   & $x^{(2)}\rightarrow$  0.8508 \\  
\hline
$N=1000$ & 0.7919 & 0.4430  & $x^{(1)}\rightarrow$  0.4724 \\ 
&    &   & $x^{(2)}\rightarrow$ 0.7662 \\  
\hline
\end{tabular}
\end{center}
\end{table}
%%%%%%%%%%%%%%%%%%%%%%%
\begin{table}[!hbt]
\begin{center}
\caption{Numerical results with $\sigma_p=10$.}
\label{TablaLast2}
\begin{tabular}{|c|c|c|c|} 
\hline
{\bf Number of tries $N$} & {\bf Acceptance Rate} & {\bf Mode-Jump Rate} & {\bf Correlation}   \\
\hline
\hline
$N=1$ (standard MH) &  0.1464 &  0.0598 & $x^{(1)}\rightarrow$  0.9097  \\ 
 &    &   & $x^{(2)}\rightarrow$ 0.9653\\ 
\hline
$N=5$  &  0.4207 &   0.2313 &  $x^{(1)}\rightarrow$  0.7536 \\ 
 &    &   & $x^{(2)}\rightarrow$ 0.8454 \\ 
\hline
$N=100$ &  0.7670 & 0.5020  &  $x^{(1)}\rightarrow$ 0.3570  \\
 &    &   & $x^{(2)}\rightarrow$  0.4607\\  
\hline
$N=1000$ & 0.8930 & 0.6520 & $x^{(1)}\rightarrow$  0.1635 \\ 
&    &   & $x^{(2)}\rightarrow$ 0.1453 \\  
\hline
\end{tabular}
\end{center}
\end{table}

From the tables, we can observe that the MTM clearly outperforms the standard MH since, as $N$ grows, the correlation decreases and the mode-jump rate increases (as does the acceptance rate) regardless of the chosen parameter $\sigma_p$ of the proposal.  
Obviously, the mode-jump rate is always less than the average value of the probability $\alpha$ of accepting a movement (the acceptance rate), since the mode-jumps represent a subset of all accepted movements. Moreover, the standard deviation $\sigma_p=10$ of the proposal pdf works better for the MTM method. In general, the MTM schemes work better with huge scaling parameters and a great-enough number of candidates $N$ (see also the discussion in the next section).

 Figures \ref{figLastIhope}(b)-(c)-(d)-(e) depict generated samples over one run. Clearly, in general we observe less than $500$ points since in certain cases a new movement is rejected and the chain remains in the same state. Namely, certain points are repeated. This effect is evident with the standard MH ($N=1$) whereas it vanishes as the number of candidates $N$ grows. Moreover, with greater $N$, the number of jumps among different modes also increases quickly. As a consequence, with the MTM technique ($N=5,100,1000$) all the features of the ``face'' (our target pdf) are completely described since the convergence of the chain is clearly speeded up.  
Therefore, with this numerical example, the main advantage of an MTM method becomes apparent: it can explore a larger portion of the sample space without a decrease of the acceptance rate, or even an increase thereof.  

%%%%%%%%%%%%%%%%%%
\section{Discussion}
%%%%%%%%%%%%%%%%%%
\label{SectConcl}

In this work, we have studied the flexibility in the design of MTM techniques. We have introduced an MTM with generic weight functions (the analytic form can be chosen arbitrarily) and different proposal densities (each candidate can be drawn from a different pdf) combining the algorithms in \citep{Casarin10} and \citep{Pandolfi10}.
 Moreover, we have proposed a general framework for construction of acceptance probabilities in the MTM schemes, providing also specific examples. Finally, we have also designed a MTM algorithm without the need of  generating randomly the reference points \citep{RobertBlog}. We have proved that the novel techniques satisfy the detailed balance condition, and carried out numerical simulations. 
Observing the theoretical workings and the numerical results, we can infer the following conclusions and observations: 
\begin{enumerate}
\item {\em General considerations:} The classical MTM method, proposed in \citep{Liu00}, clearly outperforms the standard MH algorithm using the same proposal pdf, in the sense that as the number of candidates increases, $N\rightarrow \infty$, then the correlation decreases quickly to zero (see Section \ref{IDPROP} for further considerations). If a designed MTM scheme does not fulfill this property, then it is totally useless since the computational cost increased but the performance is not improved. Suitable MTM methods can be applied efficiently to any kind of target distributions (bounded or unbounded, with heavy tails or not), as shown in our numerical simulations (see Section \ref{HeavyTail}). 
Moreover, the advantages of using an MTM technique w.r.t. a standard MH algorithm clearly grow as the dimensionality of the target increases.   

\item {\em MTM schemes as black-box algorithms:} the numerical simulations show that, with a suitable number of tries $N$, the MTM methods provide good results independently of the choice of the parameters of the proposal. 
Therefore, it is important to remark that, even if no information about the target is available (for instance, about the location of the modes), an MTM scheme allows the use of a proposal pdf with a huge scaling parameter in order to explore quickly different regions of the space. Indeed, using a great-enough number of tries, this black-box approach is quite robust and always gives satisfactory performance. On other hand, with a huge scaling parameter, a standard MH usually produces a very small rate of jumps and, as a consequence, a very high correlation. 
%This consideration appears clear in the numerical results .  

\item {\em Choice of the weights:} the possibility to choose any bounded and positive weight functions makes the MTM scheme easier to be designed since the user should not check any conditions to use suitable weights (as to check symmetry of the function $\lambda$, for instance) independently of the choice of the proposal pdfs. Namely, the proposal distribution and the weight functions can be selected separately, to fit well to the specific problem and to improve the performance of the technique. Note that, in some MTM approaches the symmetry condition of the function $\lambda$ can be complicated, see for instance \citep{LucaJesse11,Qin01}.  

Further theoretical or numerical studies are needed to determine the best choice of weight functions given a certain proposal and target density.
We find that the weights of the analytic form proposed in \citep{Liu00} (see for instance Eq. \eqref{Weightforms}) usually provide better results. Within this class, the importance weights $\omega_i(y_i)=\frac{p(y_i)}{\pi_i(y_i|x)}$, based on the importance sampling principle  \citep{Liu04b,Robert04}, appear to be a good choice in theory. Numerical results also suggest that weights simply proportional to the target density $\omega_i(y_i)=p(y_i)$ can provide good performance.
In \citep{Bedard12} the authors note that importance weights place higher probability on selecting candidates that are further away from the current
state of the chain, but finally they prefer to use weights proportional to the target density based on numerical results. 

If the evaluation of the target $p(x)$ is computationally expensive such that the target function can not be included in the calculations of the weights,  then the weight functions of the analytic class $\omega_i(y_i,x)=p(y_1)\pi_i(x|y_i)\lambda(x,y_i)$ proposed in \citep{Liu00} cannot be used. Indeed, it is impossible to find a symmetric function $\lambda(x,y)=\lambda(y,x)$ in order to remove the dependence on $p(x)$ in the weights (in this case there is just one possibility that $p(x)$ is constant, i.e., $p(x)=p(y)$ for all $x,y\in\mathcal{D}$).
In this case, a possible choice of the weights can be proportional to the proposal pdfs, namely $w(y_i)=\pi(x|y_i)$ for instance.  Clearly, it is not the optimal choice but, also in this case, the MTM can help to explore easily a larger portion of the sample space w.r.t.  standard MH (see Section \ref{DiffWeightsSect}).
\item {\em Use of different proposal pdfs:} a MTM scheme with different proposal densities can be a very powerful framework mainly to tackle applications with high dimensionality and target distributions with several modes. In our opinion, the most promising scenario is to use different independent proposal distributions updating certain parameters (as mean and variance) each iteration of the chain, or selecting the best proposal among a set of functions (see Section \ref{IDPROP} for further considerations). In this adaptive framework, the independent proposal pdfs could improved  to fit better w.r.t. the target. This scheme has not been already exploited completely. 
It is important to remark that, in order to obtain a fair comparison among the generated candidates, it is recommendable to use proposal functions with the same area below, for instance they can be normalized. 
\item {\em Flexibility of the acceptance probabilities:} we have shown there are certain freedom degrees in the design of an MTM algorithm, specifically in the choice of the acceptance probability $\alpha$. This is also confirmed by other works in literature that design suitable MTM schemes with correlated candidates but they are quite different (the strategies in \citep{LucaJesse11,Qin01} generate the candidates sequentially, whereas the approach in \citep{Craiu07} uses a block philosophy).
However, although the detailed balance condition is always satisfied in all cases, the performance is different. Numerical results suggest that $\alpha$ functions as close as possible to the standard MTM method \citep{Liu00}, using also the weights of the analytic form in Eq. \eqref{Weightforms}, perform better results. Similar considerations can be done about the standard MH algorithm \citep{Barker65,Hastings70,Peskun73}.   
\item {\em Reference points:} we have described a possible MTM algorithm without drawing reference points. As seen in the numerical results, in this case it seems to exist an optimal value of the number of candidates $N$. As $N\rightarrow \infty$ the performance becomes very poor.
 Therefore, we can figure out  that the ``secret'' of the good performance of the standard MTM scheme in \citep{Frenkel96,Liu00} is contained in the random generation of the reference points. However, there exists an important special case where the reference points are completely unnecessary: using independent proposal densities. In this case, the reference points can be set deterministically, equal to the previous generated candidates. 
This scheme, using just one proposal (drawing $N$ candidates from the same pdf) jointly with importance weights, appears as the easiest and natural procedure to combine the classical MH algorithm and importance sampling \citep{Robert04} (see Figure \ref{fig4}(a)).
\item {\em Number of candidates:} All the schemes proposed in literature and also in this work use a fixed number of candidates $N$. An important improvement would consist on tuning adaptively the number $N$ depending on the discrepancy between target and proposal distributions. To do this, a certain measure is needed, for instance, as the {\it effective sample size} of the importance sampling framework \citep{Liu04b,Robert04}. Clearly, this idea could be more effective using independent proposal pdf since it is necessary to measure the discrepancy between the proposal and the target functions (with a random walk, for instance, the mean of the proposal changes each step and the distance w.r.t. the target varies as well).
Another possibility could be to combine MTM and the delayed rejection method \citep{Mira01,Tierney99}.
With this kind of procedures, the optimal trade off between computational cost and performance would be achieved. 

%%%%%%%%%%%%%%%%%%%%%%%%%%%%%%%%%%%%%%%%%%%%%%%%
%%%%%%%%%%%%%%%%%%%%%%%%%%%%%%%%%%%%%%%%%%%%%%%%
%%%%%%%%%%%%%%%%%%%%%%%%%%%%%%%%%%%%%%%%%%%%%%%%
%%%%%%%%%%%%%%%%%%%%%%%%%%%%%%%%%%%%%%%%%%%%%%%%
%%%%% CONSIDERACION LA VARIANZA... pequena y grande de la proposal %%%%%%%%
%%%%% creo MTM mejor con varianza grande %%%%%%%%%%%%%%%%%%%%%%%%
%%%%% o funciona mejor que MH con varianza muy grande %%%%%%%%%%%%%%%%%
%%%%%%%%%%%%%%%%%%%%%%%%%%%%%%%%%%%%%%%%%%%%%%%%
%%%%%%%%%%%%%%%%%%%%%%%%%%%%%%%%%%%%%%%%%%%%%%%%
%%%%%%%%%%%%%%%%%%%%%%%%%%%%%%%%%%%%%%%%%%%%%%%%

%$N$ adaptive ...using importance sampling theory... and combination with data augmentation...
\end{enumerate}

\section{Acknowledgments}

We would like to thank the Reviewers for their comments which have helped us to improve the first version of manuscript.
Moreover, this work has been partially supported by Ministerio de Ciencia e Innovación of Spain (project MONIN, ref. TEC-2006-13514-C02- 01/TCM, Program Consolider-Ingenio 2010, ref. CSD2008- 00010 COMONSENS, and Distribuited Learning Communication and Information Processing (DEIPRO) ref. TEC2009-14504-C02-01) and Comunidad Autonoma de Madrid (project PROMULTIDIS-CM, ref. S-0505/TIC/0233).

\bibliography{bibliografia}

\begin{thebibliography}{10}

\bibitem{Barker65}
A.~A. Barker.
\newblock Monte {C}arlo calculations of the radial distribution functions for a
  proton-electron plasma.
\newblock {\em Australian Journal of Physics}, 18:119--133, 1965.

\bibitem{Bedard12}
M.~B{\'e}dard, R.~Douc, and E.~Mouline.
\newblock Scaling analysis of multiple-try {MCMC} methods.
\newblock {\em Stochastic Processes and their Applications}, 122:758--786,
  2012.

\bibitem{Brooks98}
S.~P. Brooks.
\newblock Markov {C}hain {M}onte {C}arlo method and its application.
\newblock {\em Journal of the Royal Statistical Society. Series D (The
  Statistician)}, 47(1):69--100, 1998.

\bibitem{Casarin10}
R.~Casarin, R.~V. Craiu, and F.~Leisen.
\newblock Interacting multiple try algorithms with different proposal
  distributions.
\newblock {\em Statistics and Computing}, pages 1--16, December 2011.

\bibitem{Craiu07}
R.~V. Craiu and C.~Lemieux.
\newblock Acceleration of the {M}ultiple-{T}ry {M}etropolis algorithm using
  antithetic and stratified sampling.
\newblock {\em Statistics and Computing}, 17(2):109--120, 2007.

\bibitem{Devroye86}
L.~Devroye.
\newblock {\em Non-Uniform Random Variate Generation}.
\newblock Springer, 1986.

\bibitem{Fitzgerald01}
W.~J. Fitzgerald.
\newblock {M}arkov {C}hain {M}onte {C}arlo methods with applications to signal
  processing.
\newblock {\em Signal Processing}, 81(1):3--18, January 2001.

\bibitem{Frenkel96}
D.~Frenkel and B.~Smit.
\newblock {\em Understanding molecular simulation: from algorithms to
  applications}.
\newblock Academic Press, San Diego, 1996.

\bibitem{Gamerman97bo}
D.~Gamerman.
\newblock {\em Markov Chain Monte Carlo: Stochastic Simulation for Bayesian
  Inference}.
\newblock Chapman and Hall/CRC, 1997.

\bibitem{Gilks95bo}
W.R. Gilks, S.~Richardson, and D.~Spiegelhalter.
\newblock {\em {M}arkov Chain {M}onte {C}arlo in Practice: Interdisciplinary
  Statistics}.
\newblock Taylor \& Francis, Inc., UK, 1995.

\bibitem{Haario99}
H.~Haario, E.~Saksman, and J.~Tamminen.
\newblock Adaptive proposal distribution for random walk {M}etropolis
  algorithm.
\newblock {\em Computational Statistics}, 14:375--395, 1999.

\bibitem{Haario01}
H.~Haario, E.~Saksman, and J.~Tamminen.
\newblock An adaptive {M}etropolis algorithm.
\newblock {\em Bernoulli}, 7(2):223--242, April 2001.

\bibitem{Hastings70}
W.~K. Hastings.
\newblock {M}onte {C}arlo sampling methods using {M}arkov chains and their
  applications.
\newblock {\em Biometrika}, 57(1):97--109, 1970.

\bibitem{Lan12}
S.~Lan, V.~Stathopoulosy, B.~Shahbaba, and M.~Girolami.
\newblock {L}angrangian dynamical {M}onte {C}arlo.
\newblock {\em arXiv:1211.3759v1}, November 2012.

\bibitem{Liang10}
F.~Liang, C.~Liu, and R.~Caroll.
\newblock {\em Advanced {M}arkov {C}hain {M}onte {C}arlo Methods: Learning from
  Past Samples}.
\newblock Wiley Series in Computational Statistics, England, 2010.

\bibitem{Liu04b}
J.~S. Liu.
\newblock {\em {M}onte {C}arlo Strategies in Scientific Computing}.
\newblock Springer, 2004.

\bibitem{Liu00}
J.~S. Liu, F.~Liang, and W.~H. Wong.
\newblock The {M}ultiple-{T}ry method and local optimization in {M}etropolis
  sampling.
\newblock {\em Journal of the American Statistical Association},
  95(449):121--134, March 2000.

\bibitem{LucaJesse11}
L.~Martino, Victor Pascual~Del Olmo, and Jesse Read.
\newblock A multi-point {M}etropolis scheme with generic weight functions.
\newblock {\em Statistics \& Probability Letters}, 82(7):1445--1453, 2012.

\bibitem{MartinoA2RMS}
L.~Martino, J.~Read, and D.~Luengo.
\newblock Improved adaptive rejection {M}etropolis sampling algorithms.
\newblock {\em arXiv:1205.5494v4}, 2012.

\bibitem{Metropolis53}
N.~Metropolis, A.~Rosenbluth, M.~Rosenbluth, A.~Teller, and E.~Teller.
\newblock Equations of state calculations by fast computing machines.
\newblock {\em Journal of Chemical Physics}, 21:1087--1091, 1953.

\bibitem{Mira01}
A.~Mira.
\newblock On {M}etropolis-{H}astings algorithms with delayed rejection.
\newblock {\em Metron}, 59:231--241, 2001.

\bibitem{Pandolfi10}
Silvia Pandolfi, Francesco Bartolucci, and Nial Friel.
\newblock A generalization of the {M}ultiple-try {M}etropolis algorithm for
  {B}ayesian estimation and model selection.
\newblock {\em Journal of Machine Learning Research (Workshop and Conference
  Proceedings Volume 9: AISTATS 2010)}, 9:581--588, 2010.

\bibitem{Peskun73}
P.H. Peskun.
\newblock Optimum {M}onte-{C}arlo sampling using {M}arkov chains.
\newblock {\em Biometrika}, 60(3):607--612, 1973.

\bibitem{Qin01}
Z.~S. Qin and J.~S. Liu.
\newblock {M}ulti-{P}oint {M}etropolis method with application to hybrid
  {M}onte {C}arlo.
\newblock {\em Journal of Computational Physics}, 172:827--840, 2001.

\bibitem{Robert04}
C.~P. Robert and G.~Casella.
\newblock {\em {M}onte {C}arlo Statistical Methods}.
\newblock Springer, 2004.

\bibitem{RobertBlog}
C.P. Robert.
\newblock ``{X}i' {A}n's {O}g, an attempt at bloggin... '' {B}log (by
  {C}hristian {P}. {R}obert).
\newblock {\em
  http://xianblog.wordpress.com/2012/01/23/multiple-trypoint-metropolis-algori%
thm/}, January 2012.

\bibitem{Roberts97}
G.~O. Roberts, A.~Gelman, and W.~R. Gilks.
\newblock Weak convergence and optimal scaling of random walk {M}etropolis
  algorithms.
\newblock {\em Annals of Applied Probability}, 7:110--120, 1997.

\bibitem{Storvik11}
G.~Storvik.
\newblock On the flexibility of {M}etropolis-{H}astings acceptance
  probabilities in auxiliary variable proposal generation.
\newblock {\em Scandinavian Journal of Statistics}, 38(2):342--358, February
  2011.

\bibitem{Tierney94}
L.~Tierney.
\newblock Markov chains for exploring posterior distributions.
\newblock {\em Ann. Statist.}, 33:1701--1728, 1994.

\bibitem{Tierney99}
L.~Tierney and A.~Mira.
\newblock Some adaptive {M}onte {C}arlo methods for {B}ayesian inference.
\newblock {\em Stat. Med.}, 18:2507--2515, 1999.

\bibitem{Zhang12}
Y.~Zhang and W.~Zhang.
\newblock Improved generic acceptance function for multi-point {M}etropolis
  algorithm.
\newblock {\em 2nd International Conference on Electronic and Mechanical
  Engineering and Information Technology (EMEIT-2012)}, 2012.

\end{thebibliography}

\end{document}